\documentclass[conference]{IEEEtran}
\IEEEoverridecommandlockouts
\usepackage{cite}
\usepackage{amsmath,amssymb,amsfonts}
\usepackage{algorithm}
\usepackage[noend]{algorithmic}
\usepackage{graphicx}
\usepackage{textcomp}
\usepackage{xcolor}
\usepackage{listings}
\usepackage{courier}
\def\BibTeX{{\rm B\kern-.05em{\sc i\kern-.025em b}\kern-.08em
    T\kern-.1667em\lower.7ex\hbox{E}\kern-.125emX}}

\begin{document}

\title{Pyramid: A General Framework for Distributed Similarity Search}


\author{\IEEEauthorblockN{Shiyuan Deng,\ \ Xiao Yan,\ \ Kelvin K.W. Ng,\ \ Chenyu Jiang,\ \ James Cheng}
\IEEEauthorblockA{\textit{Department of Computer Science} \\
\textit{The Chinese University of Hong Kong}\\
Shatin, Hong Kong \\
\{sydeng7, xyan, kwng6, cyjiang7, jcheng\}@cse.cuhk.edu.hk}
}


\maketitle

\begin{abstract}

Similarity search is a core component in various applications such as image matching, product recommendation and low-shot classification. However, single machine solutions are usually insufficient due to the large cardinality of modern datasets and stringent latency requirement of on-line query processing. We present Pyramid, a general and efficient framework for distributed similarity search. Pyramid supports search with popular similarity functions including Euclidean distance, angular distance and inner product. Different from existing distributed solutions that are based on KD-tree or locality sensitive hashing (LSH), Pyramid is based on Hierarchical Navigable Small World graph (HNSW), which is the state of the art similarity search algorithm on a single machine. To achieve high query processing throughput, Pyramid partitions a dataset into sub-datasets containing similar items for index building and assigns a query to only some of the sub-datasets for query processing. To provide the robustness required by production deployment, Pyramid also supports failure recovery and straggler mitigation.  Pyramid offers a set of concise API such that users can easily use Pyramid without knowing the details of distributed execution. Experiments on large-scale datasets show that Pyramid produces quality results for similarity search, achieves high query processing throughput and is robust under node failure and straggler.  
 

%

\end{abstract}

\section{Introduction}\label{sec:introduction}

Given a query $q\in\mathbb{R}^d$ and a similarity function $s(q, x)$, similarity search finds the item that is most similar to the query in a dataset $\mathcal{X}=\{x_i\in\mathbb{R}^d|i=1,\cdots,n\}$ according to the  
similarity function~\cite{wang2017similaritysurvey}. Popular similarity functions include Euclidean distance~\cite{datar2004locality}, angular distance~\cite{charikar2002angular} and inner product~\cite{shrivastava2014alsh}. In practice, it is commonly required to return the top $k$ most similar items to a query. Similarity search is a key component in a large number of applications including large-scale image search~\cite{philbin2007imagesearch}, semi-supervised low-shot classification~\cite{douze2018lowshot}, recommendation based on user and item embeddings~\cite{koren2009matrix}, sequence matching~\cite{berlin2015}, entity resolution~\cite{hoffart2012kore}, memory network training~\cite{chandar2016memory} and reinforcement learning~\cite{jun2017reinforcement}. Similarity search that returns the exact top $k$ neighbors is usually too costly and approximate similarity search, which returns a good portion of the exact top $k$ neighbors, suffices for most applications. Therefore, we focus on approximate similarity search in this paper.

\vspace{2mm}

\noindent \textbf{Similarity search algorithms.} Due to the importance of similarity search, many algorithms have been proposed to solve it efficiently. Existing similarity search algorithms can be roughly classified into four categories, i.e., tree-based methods~\cite{fukunage1975branch, jagadish2005idistance, ram2012conetree}, locality sensitive hashing (LSH) based methods~\cite{indyk1998approximate, datar2004locality, neyshabur2014symmetric}, vector quantization based methods~\cite{jegou2010pq, ge2013opq, babenko2014imi} and proximity graph based methods~\cite{hajebi2011knngraph, harwood2016fanng, malkov2018hnsw}. Among them, the proximity graph based methods were shown to provide the best recall-time performance\footnote{Recall-time performance measures the recall achieved within a given query processing time and higher recall indicates better performance.} in a number of empirical studies~\cite{malkov2018hnsw, wang2013graphimi, fu2019nsg}. In a proximity graph, each item is connected to a small set of items that are most similar to it in the dataset and the graph-based methods usually conduct similarity search by a walk on the proximity graph. With the proximity graph, the graph-based methods can model the fine-grained neighboring relation among items and avoid checking dissimilar items for query processing, which explains their good performance~\cite{fu2019nsg}. Among the proximity graph based methods, the Hierarchical Navigable Small World graph (HNSW)~\cite{malkov2018hnsw} represents the state-of-the-art method because of its fast index construction and good search performance. We will give a detailed introduction to HNSW in Section~\ref{sec:background}.

There are also some distributed similarity search solutions designed to handle large datasets~\cite{muja2014flann, sundaram2013streaming, zhang2016shuffle}. However, these solutions use either tree-based methods~\cite{muja2014flann} or LSH-based methods~\cite{sundaram2013streaming, zhang2016shuffle}, and scalable solutions for the more recently proposed proximity graph based methods are still lacking.  Our work proposes a distributed solution based on HNSW, the state of the art similarity search method on a single machine, to improve the performance of distributed similarity search.

\vspace{2mm}

\noindent \textbf{Single machine solutions.} \ The main challenge of similarity search on large-scale datasets is the memory required to store the raw data and index data structures, which could easily exceed the capacity of a single machine. For example, the SIFT1B dataset~\cite{jegou2011sift}, which contains 1 billion 128-dimensional SIFT descriptors of images, takes up 512GB of memory for holding the raw data if each feature is stored as a floating point number. In addition, the proximity graph of HNSW also takes a large amount of memory (often comparable with the size of the raw data) as it needs to maintain a neighbor list for every item. 

To reduce memory consumption, existing single machine solutions such as FAISS~\cite{johnson2017faiss} and Link\&Code~\cite{douze2018link} use vector quantization techniques (e.g., PQ~\cite{jegou2010pq} and OPQ~\cite{ ge2013opq}) to compress the items. However, the quantization error introduced by the compression process often harms the quality of the search results. For example, using OPQ with 8 codebooks, FAISS achieves a precision of only 25.15\% when $2^{22}$ items are probed for top-10 Euclidean nearest neighbor search\footnote{To calculate precision, the items are ranked according to their approximate similarity scores, i.e., $s(\tilde{x}, q)$. If $k'$ of the ground truth top $k$ Euclidean nearest neighbors are identified in the items ranking top $k$, the precision is $k'/k$.}. This is because, given that two items $x$ and $y$ have similarity scores $s(x, q)>s(y,q)$ with a query $q$, a small quantization error could lead to $s(\tilde{x}, q)<s(\tilde{y}, q)$ if the compressed approximations (i.e., $\tilde{x}$ and $\tilde{y}$) are used to evaluate the similarity function.


Therefore, existing single machine solutions are inadequate for large-scale similarity search as they cannot provide high quality results, which is particularly critical in applications such as e-commerce and advertising~\cite{abuzaid2019maximus}. To provide high quality search results and scale to even larger datasets (e.g., with trillions of items) we may encounter in the future, it is necessary to develop distributed solutions that can store and process queries with uncompressed data. 

\vspace{2mm}

\noindent \textbf{Requirements.} Apart from producing high quality results, a similarity search framework needs to fulfill three additional requirements for production use, i.e.,~\textit{high query processing throughput, low query processing latency, and good robustness}. Query processing throughput is the number of queries the framework can handle in unit time.  Query processing latency measures time taken to process a query and online applications typically require a query processing latency in the order of several milliseconds~\cite{baranchuk2018revisiting}. As node failures and stragglers are common in a distributed computing, the performance of the framework should also be robust under these adversarial scenarios.

\vspace{2mm}

\noindent \textbf{Our solution.} Pyramid is a distributed solution based on HNSW and supports popular similarity functions including Euclidean distance, angular distance and inner product. A naive distributed solution with HNSW is to randomly partition the dataset over the machines and build an HNSW on each machine. However, its query processing throughput is low as partitions in every machine need to be searched for processing a query. Pyramid solves the deficiency of the naive solution with novel dataset partitioning and query assignment strategies. The key idea is to build a much smaller meta-HNSW that captures the structure of the entire dataset, which allows us to build two levels of indexes and pinpoint the neighbors of a query efficiently. Specifically, by partitioning the bottom layer of the meta-HNSW, Pyramid assigns dataset items to sub-datasets of roughly equal size and ensures that items in the same sub-dataset are similar to each other. By searching the meta-HNSW, Pyramid quickly identifies the sub-datasets that are likely to contain the neighbors of a query and involves only these sub-datasets in query processing without hurting the quality of the search results. For failure recovery and straggler mitigation, Pyramid replicates the sub-datasets and their HNSWs across the machines.

We implement the index building component of Pyramid with customized code for efficient distributed execution. For query processing, we employ Zookeeper\footnote{https://zookeeper.apache.org/} to monitor the system and perform automatic failure recovery. Kafka\footnote{https://kafka.apache.org/} is used to dispatch queries to the machines and automatically handle load balancing and fault tolerance for the message queues. At the top level, we provide a set of simple and expressive high-level API to hide the low-level execution details from users.  

We tested Pyramid on three large-scale datasets, Deep500M, SIFT500M and Tiny10M. The results show that Pyramid provides high quality search results and the precision can easily reach 90\% for top 10 Euclidean nearest neighbor search. The throughput of Pyramid is over 2x compared with a naive solution that randomly partitions a dataset among the machines and builds an HNSW for each machine. Comparing with the famous FLANN library~\cite{muja2014flann} that uses tree based method for distributed similarity search, Pyramid provides a throughput that is over 100x higher and achieves better precision for the search results. Pyramid is able to keep the query processing latency within 2-3ms. Pyramid is also robust to stragglers and node failures with its replication strategy.             

\vspace{2mm}

\noindent \textbf{Paper organization.} \ The remainder of the paper is organized as follows. Section II introduces the basics about HNSW to as a background. The index building and query processing algorithms of Pyramid are discussed in Section III. Section IV introduces Pyramid's API and the designs for straggler mitigation and failure recovery. Section V presents the experiment results and Section VI surveys the related work. The concluding remarks are given in Section VII.     
\section{Hierarchical Navigable Small World Graph}\label{sec:background}

\begin{figure}
	\centering
	\includegraphics[width=0.6\linewidth]{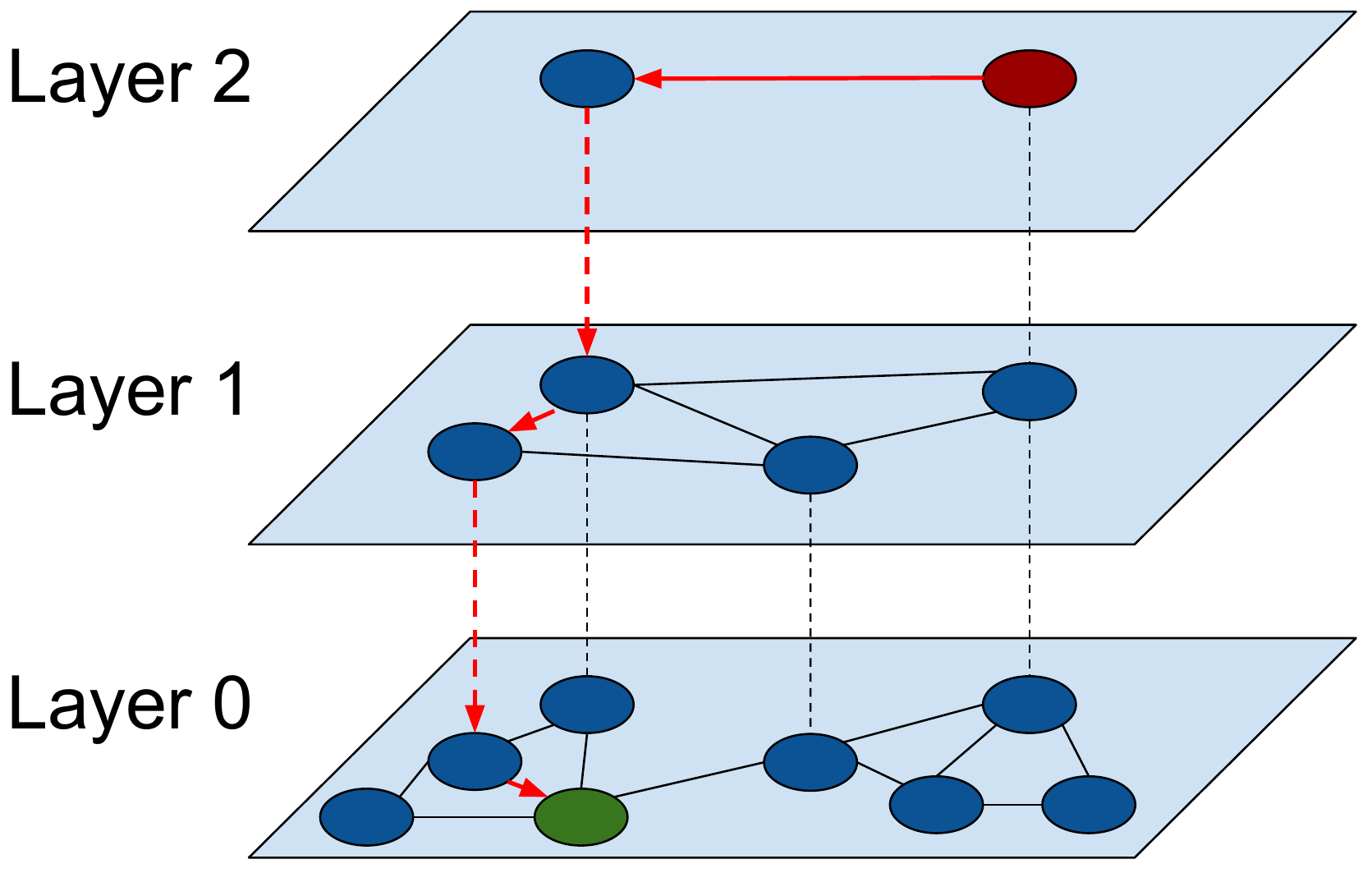}
	\caption{An illustration of an HNSW Graph}
	\label{fig:HNSW}
\end{figure}

We first introduce the query processing and graph construction procedures of Hierarchical Navigable Small World graph (HNSW)~\cite{malkov2018hnsw}, as Pyramid is a distributed solution based on HNSW. For simplicity, we omit some details in the algorithms and readers may refer to~\cite{malkov2018hnsw} for the complete algorithms. We also present the algorithms using a similarity function $s(q,x)$ instead of a distance function and a larger similarity value indicates that two items are more similar. To search for neighbors with small distances, the similarity function can be defined as negative distance. For example, $s(q,x)=-\Vert q-x \Vert$ can be used for Euclidean distance nearest neighbor search (Euclidean NNS).

The HNSW proximity graph has multiple layers as illustrated in Figure~\ref{fig:HNSW}. The bottom layer (layer 0) contains all items in the dataset, while the items in each upper layer are sampled uniformly from its previous layer. Therefore, the number of items reduces as the layer increases. Each layer of HNSW is an approximation of the $k$-nearest neighbor graph (KNN graph)~\cite{hajebi2011knngraph} and an item is connected to its approximate top $M$ neighbors in that layer, where $M$ is a user-specified number that controls the size of the graph. 

The query processing procedure of HNSW is shown in Algorithm~\ref{alg:HNSW query}, in which $C$ is the candidate queue and $W$ keeps the best results encountered so far. The graph walk starts at a fixed entry vertex at the top layer. Before reaching the bottom layer, graph walk is conducted with a search factor (that controls the size of $W$) of 1, which is also called greedy graph walk (without backtracking). Greedy graph walk moves to the best neighbor (most similar to the query) of the current vertex at every step and stops when the current vertex is more similar to the query than all its neighbors. The stopping vertex in an upper layer is used as the starting vertex for graph walk in the layer below it. For the bottom layer, graph walk is usually conducted with a search factor $l>1$, which is similar to beam search with backtracking and makes the walk less likely to be trapped at local optimal.

HNSW is an improvement of the KNN graph, which contains a single layer and each item is connected to its top neighbors. In a KNN graph, graph walk can only take small steps as the connections are local. If the starting vertex is far from the neighborhood of the query, graph walk needs a large number of steps to reach the true neighbors, which harms performance. 
The upper layers of HNSW contain uniformly sampled items from the dataset and they allow graph walk to take large steps and quickly approach the neighborhood of the query. Using a large search factor, graph walk on the bottom layer allows good exploration of the neighborhood of the query to identify the true neighbors.
       
\begin{algorithm}
	\caption{HNSW: Query Processing via Graph Walk}
	\label{alg:HNSW query}
	\begin{algorithmic}[1]
		\STATE {\bfseries Input:} HNSW graph $G$, similarity function $s$, query $q$, search factor $l$ for the bottom layer
		\STATE {\bfseries Output:} $k$ approximate neighbors of $q$
		\STATE Set queue $C.add(G.entryVertex, s(q, G.entryVertex))$
		\FOR {$t=G.maxLayer$ downto 1}
		\STATE   $C \gets$ Search-Level($G$, $t$, $q$, $s$, $C$, 1)
		\STATE	 $C \gets \{max(C)\}$
		\ENDFOR
		\STATE $C \gets$ Search-Level($G$, $0$, $q$, $s$, $C$, $l$)
		\RETURN Top $k$ vertexes with the maximum $s(q, x)$ in $C$ 
		\\
		\mbox{}
		\STATE \textbf{function} Search-Level($G$, $level$, $q$,  $s$, $C$, $factor$)
		\begin{ALC@g}
		\STATE $W=C$
		\WHILE{$C.size \! >\! 0$ \& $s(q, max(C))\!\ge \! s(q, min(W))$}
			\STATE $v_{curr} \gets C.pop\_max()$
			\FOR {each neigbour $v$ of $v_{curr}$ in $G.level$}
				\IF {$v$ is not checked}
					\STATE $C.add(v, s(q, v))$, $W.add(v, s(q, v))$
				\ENDIF				
			\ENDFOR
			\STATE Resize $W$ to $factor$ by keeping the similar items  
         \ENDWHILE
         \RETURN $W$
		\end{ALC@g}
	\end{algorithmic}
\end{algorithm} 
The graph construction procedure of HNSW is shown in Algorithm~\ref{alg:HNSW graph construction}. The items in the dataset are inserted sequentially into the graph to build the HNSW. For an item $x$, the highest layer (denoted by $u$ in the algorithm) it can appear is first generated by an exponential distribution. Then graph search is conducted using $x$ as the query and the only difference from Algorithm~\ref{alg:HNSW query} is that large search factor ($l$) is used for all layers below $u+1$. In each of these layers, graph search is used to find the top $M$ neighbors of $x$ and $x$ is connected to them using directed edges.

\begin{algorithm}
	\caption{HNSW: Graph Construction}
	\label{alg:HNSW graph construction}
	\begin{algorithmic}[1]
		\STATE {\bfseries Input:} dataset $\mathcal{X}$, similarity function $s$, maximum vertex degree $M$, search factor $l$ for graph walk 
		\STATE {\bfseries Output:} HNSW proximity graph $G$ 
		\STATE Initialize $G=\emptyset$
		\FOR {each item $x$ in $\mathcal{X}$ }
		\STATE $u=random()$ and $C.add(G.entryVertex)$
		\FOR {$t=G.maxLayer$ downto $u+1$}
		\STATE   $C \gets$ Search-Level($G$, $t$, $x$, $s$, $C$, 1)
		\STATE	 $C \gets \{max(C)\}$
		\ENDFOR
		\FOR {$t=u$ downto $0$}
		\STATE   $C \gets$ Search-Level($G$, $t$, $x$, $s$, $C$, $l$)
		\STATE	 Connect $x$ to its top $M$ neighbors in $C$ for graph $G.t$
		\STATE	 $C \gets \{max(C)\}$
		\ENDFOR
		\ENDFOR
		\RETURN $G$ 
	\end{algorithmic}
\end{algorithm}  

In a number of empirical studies, HNSW is found to significantly outperform other similarity search algorithms, including tree-based methods, LSH-based methods and vector quantization based methods. It is reported that the search complexity of HNSW scales with $O(\log n)$ with $n$ being the cardinality of the dataset~\cite{malkov2018hnsw}, which is favorable for large datasets. Index construction of HNSW is also efficient as it does not require to build an exact KNN graph at each layer. Although HNSW is originally designed for metric similarity functions such as Euclidean distance and edit distance, it has been shown recently that HNSW also achieves the state-of-the-art performance for maximum inner product search (MIPS)~\cite{morozov2018mips}. Therefore, the efficiency in index building, excellent similarity search performance and generality to similarity functions make HNSW an ideal choice for building a distributed similarity search solution based on it.    

\section{Pyramid} \label{sec:design}

In this section, we introduce the algorithmic aspect of Pyramid, i.e., the index building and query processing procedures. We also discuss the special considerations in Pyramid for MIPS.   

First, we motivate the design of Pyramid by analyzing the deficiency of a naive solution. For distributed similarity search with HNSW, a straightforward solution (denote as HNSW-naive) is to randomly partition a dataset among workers in a cluster and build an independent HNSW graph on each worker. A query $q$ is distributed to all workers and each worker processes the query with its own HNSW graph using Algorithm~\ref{alg:HNSW query}. The final search results are obtained by merging and re-ranking the partial results reported by the workers. However, a query invokes computation on all workers in HNSW-naive, which results in low query processing throughput. 

If a query is handled by only some rather than all of the workers, query processing throughput can be improved as each query invokes less workload. This is the main motivation of Pyramid's design, which is achieved by dataset partitioning and query assignment. In the index building phase, Pyramid partitions the dataset into sub-datasets containing items similar to each other and assigns each sub-dataset to a worker. Due to the partitioning, some sub-datasets are likely to contain the neighbors for a query while others are not. Then for query processing, it is sufficient to handle the query by the workers holding these potential sub-datasets and the other workers do not need to be involved, which results in high query processing throughput. In Pyramid, both dataset partitioning and query assignment are conducted with a small meta-HNSW built on samples from the dataset.

\subsection{Index Building}

The index building procedure of Pyramid is shown in Algorithm~\ref{alg:Pyramid Index}. As the original dataset $\mathcal{X}$ may be very large, we first sample a small dataset $\mathcal{X}'$ with size $n'$ from it. Then, kmeans with $m$ centers is conducted on the sample dataset $\mathcal{X}'$ and we assign a weight to each kmeans center, which is set as the number of items it has from $\mathcal{X}'$. The meta-HNSW $G_m$ is built on these kmeans centers using Algorithm~\ref{alg:HNSW graph construction}. We partition the bottom layer (which is a proximity graph) of $G_m$ into $w$ balanced graph partitions (in the sense that each graph partition has similar total vertex weights) and try to minimize the number of edges across the graph partitions. Note that by minimizing the number of cross edges, we ensure that items in each graph partition are similar to each other. Currently, we use the Karlsruhe Fast Flow Partitioner algorithm~\cite{sanders2012think} for partitioning, which adopts an efficient multi-level local improvement strategy to search for the best partitioning. Finally, for each item $x$ in the original dataset $\mathcal{X}$, we assign it to sub-dataset $\mathcal{X}^i$ according to its most similar item in $G_m$. For each sub-dataset $\mathcal{X}^i$, we build a sub-HNSW $G_i$ independently. Note that there is a one to one mapping between the sub-datasets and the graph partitions (of the bottom layer of $G_m$) and the $i$-th partition corresponds to sub-dataset $\mathcal{X}^i$.

The design of Algorithm~\ref{alg:Pyramid Index} is a joint consideration of~\textit{efficiency, load balancing and statistical stability}. As the original dataset is large, we use a smaller sample dataset $\mathcal{X}'$ as its surrogate to speed up the index building process. Assume that each item in $\mathcal{X}$ is equally likely to be accessed by queries, the sub-datasets $\mathcal{X}^1,\mathcal{X}^2,\cdots,\mathcal{X}^w$ should have roughly equal size to balance their workloads. Therefore, we set the weight of a vertex in $G_m$ as the number of items it has from $\mathcal{X}'$ and ensure that the partitions of $G_m$ have similar total vertex weights. There may be scenarios that some items in $\mathcal{X}$ are hot (more likely to be accessed by queries) and we are given a set of sample queries. In this case, we can set the weight of a vertex in $G_m$ as the frequency it appears in the top $k$ similarity search results of the queries for load balancing. We do not directly sample $m$ items from the dataset to build the meta-HNSW as $m$ may be small and a small sample may not reflect the distribution of the entire dataset. Empirically, we observed that a small meta-HNSW is already sufficient for good performance and a large meta-HNSW is not favorable as it prolongs the query processing time. Therefore, we conduct kmeans on a larger sample (i.e., $\mathcal{X}'$ with $n'$ items) of the dataset to obtain the vertexes in the meta-HNSW for statistical stability.

\begin{algorithm}
	\caption{Pyramid: Index Construction}
	\label{alg:Pyramid Index}
	\begin{algorithmic}[1]
		\STATE {\bfseries Input:} Dataset $\mathcal{X}$, sample size $n'$, meta-HNSW size $m$, number of sub-HNSWs $w$
		\STATE {\bfseries Output:} A meta-HNSW $G_m$ and $w$ sub-HNSWs
		\STATE Randomly sample $n'$ items from dataset $\mathcal{X}$ to form $\mathcal{X}'$ 
		\STATE Conduct kmeans with $m$ centers on $\mathcal{X}'$
		\STATE Build the meta-HNSW $G_m$ on the kmeans centers 
		\STATE Partition the bottom layer graph of $G_m$ into $w$ partitions 
		\FOR {each item $x$ in the original dataset $\mathcal{X}$}
		\STATE Find its top neighbor $n(x)$ in $G_m$ with Algorithm~\ref{alg:HNSW query}
		\IF {$n(x)$ is in the $i$-th graph partition of $G_m$} 
		\STATE Assign $x$ to sub-dataset $\mathcal{X}^i$ 
		\ENDIF
		\ENDFOR
		\FOR{sub-dataset from $1$ to $w$ }
		\STATE Build sub-HNSW $G_i$ on sub-dataset $\mathcal{X}^i$
		\ENDFOR
	\end{algorithmic}
\end{algorithm} 

\vspace{2mm}
\noindent \textbf{Distributed workflow.} The index construction procedure starts with each worker reading a part of the dataset from the distributed file system. Then each worker samples some items from its local dataset (according to the cardinality of the entire dataset $\mathcal{X}$ and its local dataset, and the total sample size $n'$) and the workers conduct distributed kmeans with $m$ centers together. The centers are kept in one of the workers for meta-HNSW construction and graph partitioning. When graph partitioning finishes, the worker also builds a one-to-one mapping between the graph partitions (and the sub-dataset indexes) and the workers. After that, the meta-HNSW is broadcast to all workers along with related data structures. The workers decide for each item in its local dataset which sub-dataset it belongs to using the meta-HNSW and shuffles the items to their destination workers in parallel. When data shuffle finishes, each worker builds an HNSW on its own sub-dataset.        

\subsection{Query Processing}

The query processing procedure of Pyramid is shown in Algorithm~\ref{alg:Pyramid Query}. Query processing starts by finding the top $K$ neighbors of the query in the meta-HNSW $G_m$, which can be done very efficiently as the meta-HNSW is usually very small. For each graph partition of the bottom layer of $G_m$, if it contains one or more of these $K$ neighbors, the query will be dispatched to its corresponding sub-dataset. We call $K$ the branching factor as it controls how many sub-datasets a query will be forwarded to and a larger $K$ means more sub-datasets will be involved. For each sub-dataset that receives the query $q$, HNSW-based graph walk is conducted to find the top $k$ neighbors of $q$. The final search results are obtained by selecting the top $k$ neighbors from the partial results returned by the sub-datasets. The query processing algorithm ensures that only some of the sub-datasets are activated for a query, which helps to achieve high query processing throughput.         

\begin{algorithm}
	\caption{Pyramid: Query Processing with meta-HNSW}
	\label{alg:Pyramid Query}
	\begin{algorithmic}[1]
		\STATE {\bfseries Input:} Query $q$, branching factor $K$, number of required neighbors $k$, meta-HNSW $G_m$, $w$ sub-HNSWs 
		\STATE {\bfseries Output:} $k$ approximate neighbors of $q$
		\STATE Initialize $resSet=\emptyset$
		\STATE Find the top $K$ neighbors of $q$ in $G_m$ with Algorithm~\ref{alg:HNSW query} 
		\FOR {the $w$ graph partitions of the bottom layer of $G_m$}
		\IF {partition $i$ contains some of the $K$ neighbors}  
		\STATE Search sub-HNSW $G_i$ for $k$ neighbors of $q$
		\STATE Add the results to $resSet$
		\ENDIF
		\ENDFOR 
		\STATE Extract the top $k$ neighbors of $q$ from $resSet$  
	\end{algorithmic}
\end{algorithm} 

\vspace{2mm}
\noindent \textbf{Distributed workflow.} When a query comes, it is assigned to a random worker as the coordinator. The coordinator searches the meta-HNSW with the query and decides the sub-datasets that will be involved in processing the query according to the search results. Then, the query is dispatched to the corresponding workers and the workers search their own HNSW with the query. Each involved worker returns $k$ tuples of~\textit{(item id, similarity score)} to the coordinator. When all responses for a query are gathered, the coordinator selects items with the top $k$ similarity scores as the final results. 

\begin{figure}
\centering
\includegraphics[width=0.8\linewidth]{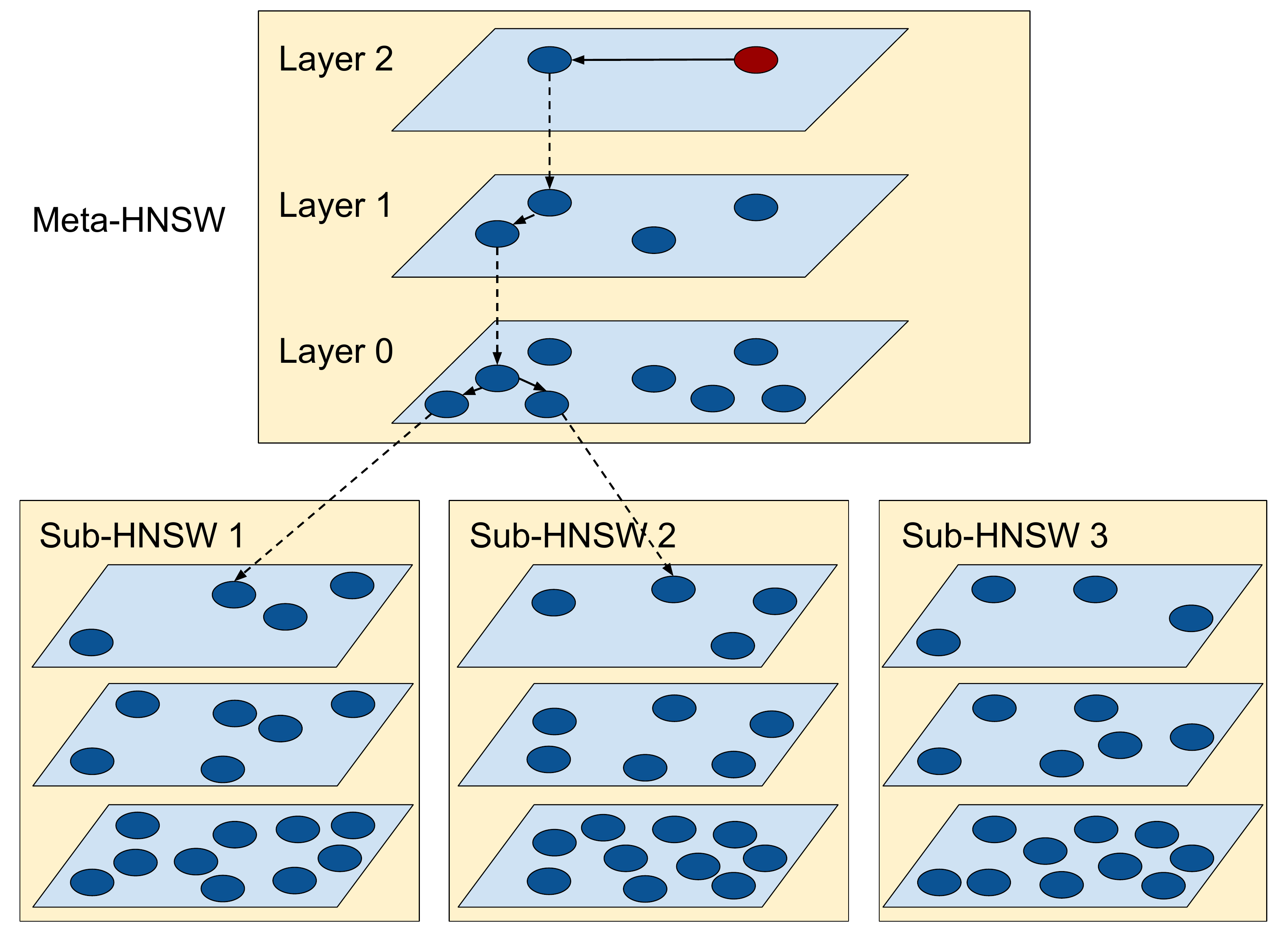}
\caption{An Illustration of Pyramid}
\label{fig:Pyramid}
\end{figure}

We give an illustration of Pyramid in Figure~\ref{fig:Pyramid}. If we treat the sub-HNSWs on the workers as the bottom layer, the meta-HNSW is equivalent to some common upper layers of the sub-HNSWs. By searching the meta-HNSW, we can quickly identify the sub-HNSWs that are likely to contain the neighbors of a query. This is analogous to the upper layers of an HNSW, which help quickly approach the neighborhood of the query. By using the bottom layer of the meta-HNSW for dataset partitioning, we ensure that items in the same sub-dataset are similar to each other.

\subsection{The Generality of Pyramid}

One important design goal of Pyramid is generality, which means that Pyramid should work for search with popular similarity functions including Euclidean distance, angular distance and inner product. As HNSW is originally designed for Euclidean distance, Algorithm~\ref{alg:Pyramid Index} and Algorithm~\ref{alg:Pyramid Query} naturally work for Euclidean NNS by using $s(q,x)=-\Vert q-x \Vert$ as the similarity function. Angular distance (i.e., $d(q, x)=\frac{q^{\top}x}{\Vert q \Vert \Vert x \Vert}$) is monotone to Euclidean distance when both query and item have unit norm, which suggests that we can transform angular similarity search into Euclidean NNS. By normalizing the items to unit norm before index construction and normalizing the query before query processing, Algorithm~\ref{alg:Pyramid Index} and Algorithm~\ref{alg:Pyramid Query} also work for angular distance. Although MIPS can also be transformed into Euclidean NNS, it has been shown that the transformation harms the performance of proximity graph based methods~\cite{morozov2018mips}. In the following discussion, we also analyze the problems of directly using Algorithm~\ref{alg:Pyramid Index} and Algorithm~\ref{alg:Pyramid Query} for MIPS and present our solutions.

\begin{figure}
	\centering
	\includegraphics[width=0.7\linewidth]{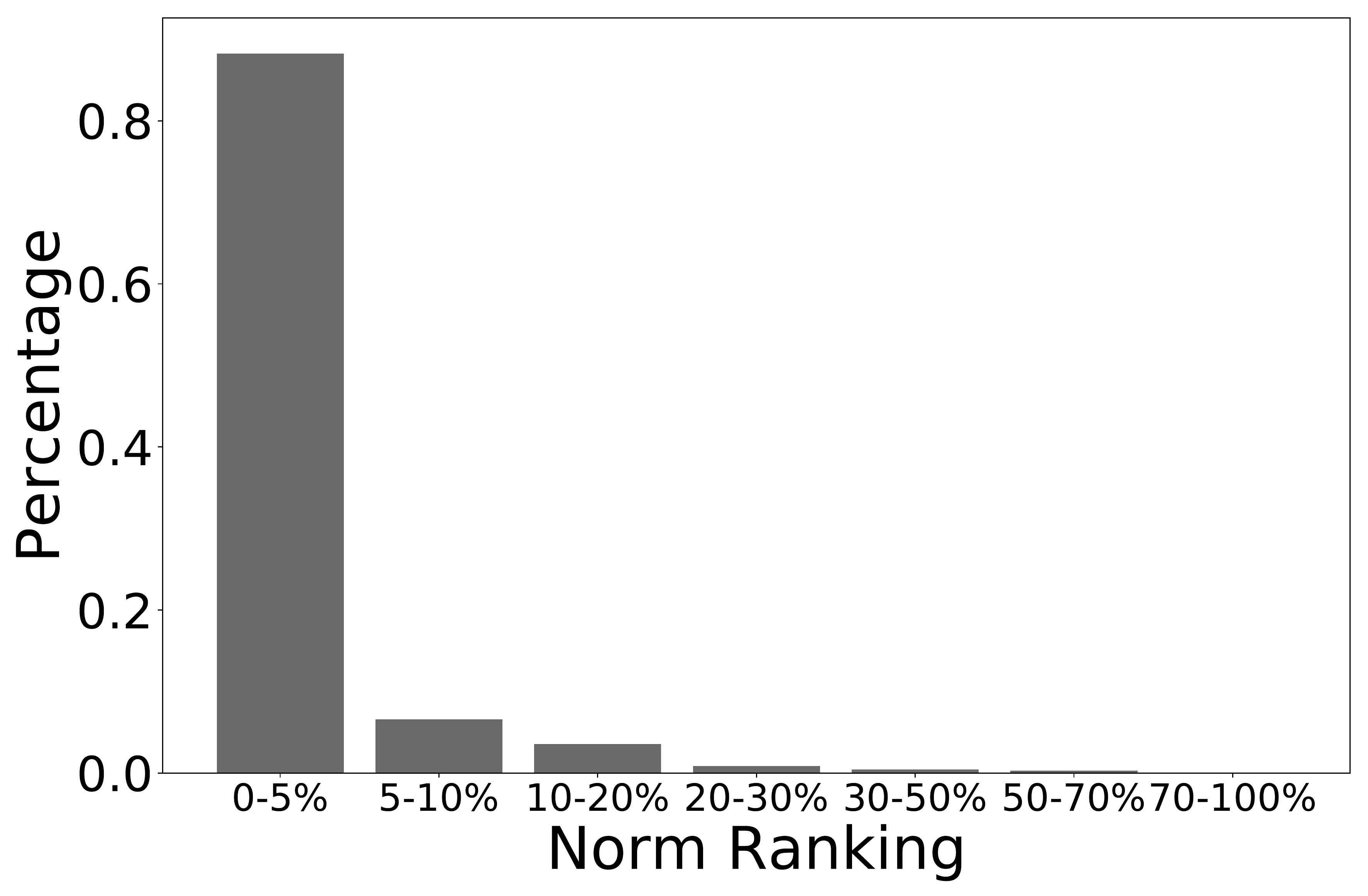}
	\caption{Result distribution for MIPS}
	\label{fig:MIPS demo}
\end{figure}

\vspace{2mm}
\noindent \textbf{Pyramid for MIPS.} \ One interesting property of MIPS is that items with large norm are very likely to be the search results, which we show in Figure~\ref{fig:MIPS demo}. For the ImageNet dataset, which contains about 2 million, 150 dimensional descriptors of images, we found the exact top-10 MIPS results of 1,000 randomly selected queries with linear scan. This gives us a result set containing 10,000 items (without deduplication) and we calculated the percentage that items with different norm percentiles take in this result set. For example, the first bar (from left to right) in Figure~\ref{fig:MIPS demo} means that items ranking top 5\% in norm takes up 93.1\% of the result set. This phenomenon can be explained by the fact that high dimensional vectors tend to have a large angle with each other. Thus, the influence of norm on inner product is decisive since $q^{\top}x=\Vert q \Vert \Vert x \Vert \cos(\theta_{q,x})$, in which $\theta_{q,x}$ is the angle between $q$ and $x$. This is contrasted with Euclidean NNS, for which each item should have equal probability to be the search result if the query comes from the same distribution of the dataset items. 

The bias towards items with large norm in MIPS causes problems for both dataset partitioning and query processing. Items with large norm tend to be strongly connected with each other in the inner product HNSW as they are likely to be the results of MIPS. In Algorithm~\ref{alg:Pyramid Index}, we partition the meta-HNSW by minimizing the number of cross edges, which means the large norm items will be put into the same graph partition (denote this partition as the large norm partition). For dataset partitioning, most items will find their MIPS in the large norm partition and this means that one of the sub-datasets will be much larger than the others. This can cause the worker holding the large sub-dataset to run out of memory. For query processing, the larger norm partition is very likely to contain the top-$K$ MIPS of most queries for meta-HNSW search, which makes the worker holding the large norm partition much more heavily loaded than the other workers and may become a straggler in the system.

\begin{algorithm}
	\caption{Pyramid: Index Construction for MIPS}
	\label{alg:Pyramid Index MIPS}
	\begin{algorithmic}[1]
		\STATE {\bfseries Input:} Dataset $\mathcal{X}$, sample size $n'$, meta-HNSW size $m$, number of sub-HNSWs $w$, replication factor $r$
		\STATE {\bfseries Output:} A meta-HNSW $G_m$ and $w$ sub-HNSWs
		\STATE Randomly sample $n'$ items from dataset $\mathcal{X}$ 
		\STATE Normalize the sampled items to unit norm to form $\mathcal{X}'$   
		\STATE Conduct spherical kmeans with $m$ centers on $\mathcal{X}'$
		\STATE Build the meta-HNSW $G_m$ on the kmeans centers 
		\STATE Partition the bottom layer graph of $G_m$ into $w$ partitions 
		\FOR {each item $x$ in the original dataset $\mathcal{X}$}
		\STATE Find its top neighbor $n(x)$ in $G_m$ with Algorithm~\ref{alg:HNSW query}
		\IF {$n(x)$ is in the $i$-th graph partition of $G_m$} 
		\STATE Assign $x$ to sub-dataset $\mathcal{X}^i$
		\ENDIF
		\ENDFOR
		\FOR{each of the $w$ graph partitions of $G_m$}
		\FOR{each vector in the $i$ graph partition }
		\STATE Find its top $r$ MIPS neighbor in dataset $\mathcal{X}$
		\STATE Add these items to sub-dataset $\mathcal{X}^i$ 
		\ENDFOR  
		\ENDFOR
		\FOR{sub-dataset from $1$ to $w$ }
		\STATE Build sub-HNSW $G_i$ on sub-dataset $\mathcal{X}^i$
		\ENDFOR
	\end{algorithmic}
\end{algorithm}    

To solve the aforementioned problems, we use Algorithm~\ref{alg:Pyramid Index MIPS} to build index for MIPS. Note that Algorithm~\ref{alg:Pyramid Index MIPS} uses inner product as similarity function for all operations.  There are two main changes in Algorithm~\ref{alg:Pyramid Index MIPS} compared with Algorithm~\ref{alg:Pyramid Index}. Firstly, the sampled items are normalized to unit norm in line 4 and the kmeans is spherical kmeans~\cite{auvolat2015clustering} in line 5, which ensures that the centers have unit norm. Therefore, all items in the meta-HNSW have unit norm and each graph partition contains vectors pointing to similar directions. In line 8-11, each item in the original dataset $\mathcal{X}$ is assigned to the graph partition that contains its MIPS. As all meta-HNSW vectors have unit norm, the MIPS of an item is also most similar to it in direction. Thus, each sub-dataset $\mathcal{X}^i$ covers items pointing to similar direction, which avoids the problem that the large norm partition attracts much more items than the others. The problem that the large norm partition is hot for query processing is also solved as the query is assigned to sub-datasets that are similar to it in direction and no sub-dataset is more likely to attract queries. 

In lines 12-15, Algorithm~\ref{alg:Pyramid Index MIPS} introduces an additional item assignment stage compared with Algorithm~\ref{alg:Pyramid Index} and the motivation is to reduce the number of sub-datasets needed to process a query. We have already assigned items pointing to similar directions to the same sub-dataset. However, items with large norm are likely to be the result of MIPS and it is possible that these items are scattered across many sub-datasets. This means that the query needs to access all these sub-datasets to achieve high recall, which hurts query processing throughput. In lines 12-15, we add the top $r$ MIPS neighbors of each meta-HNSW vector to its corresponding sub-datasets, which enables items with large norm to be assigned to multiple sub-datasets. Empirically, we found that setting $mr$ to a relatively small fraction (e.g., several percent) of the dataset cardinality $n$ already leads to good performance, which means that the memory overhead of this allocation is small. Finding the top $r$ MIPS neighbors of an vector in the original dataset $\mathcal{X}$ can be done approximately and efficiently using LSH-based methods~\cite{neyshabur2014symmetric,shrivastava2014alsh}. The query processing procedure for MIPS is exactly the same as Algorithm~\ref{alg:Pyramid Query}.   

\section{System Design and Implementation}

In this section, we introduce the system architecture and API of Pyramid. The fault tolerance and straggler mitigation strategies of Pyramid are also discussed. We will make Pyramid open source.       

\begin{figure}
	\centering
	\includegraphics[width=0.6\linewidth]{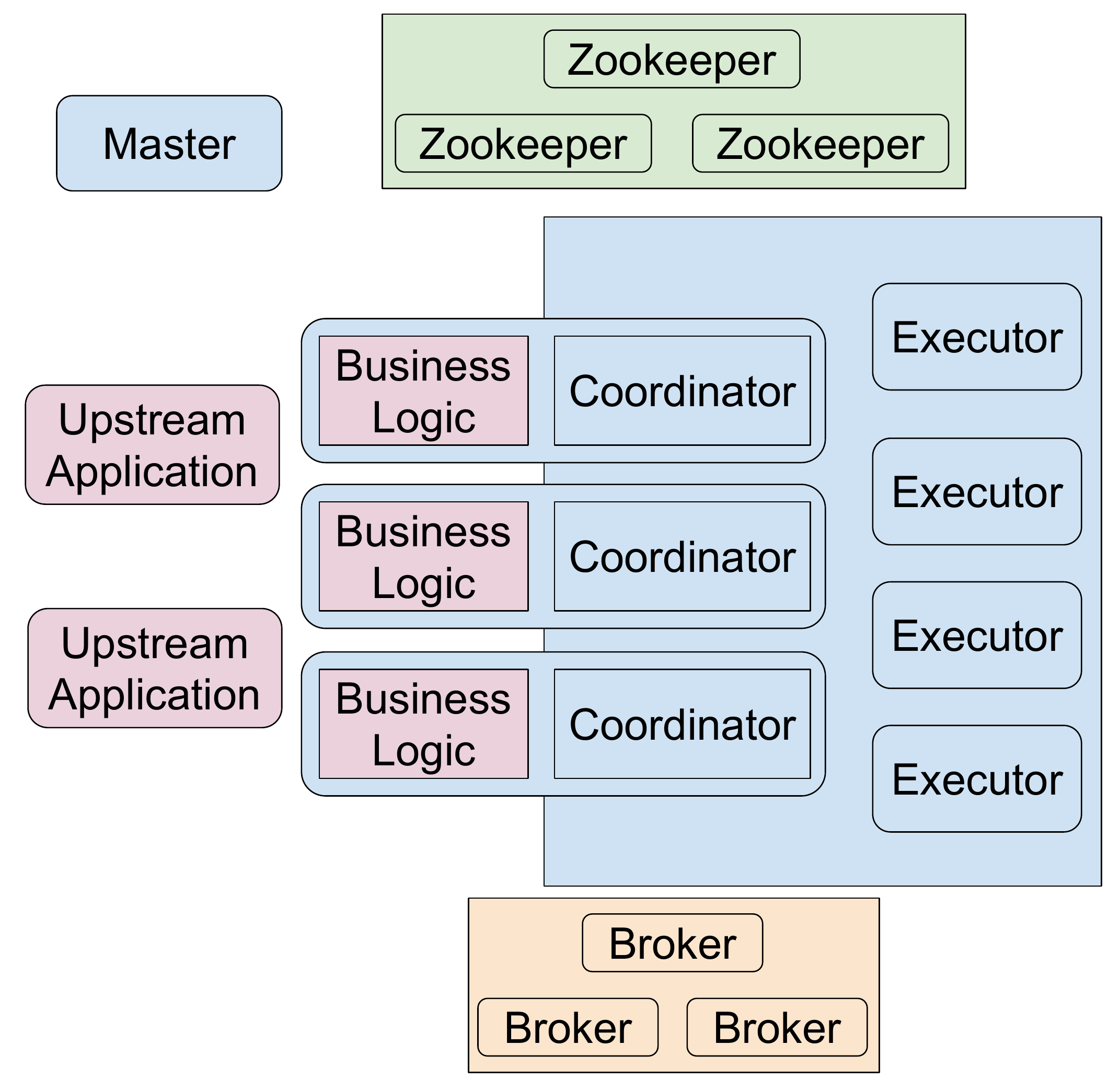}
	\caption{System Architecture of Pyramid}
	\label{fig:sys_arch}
\end{figure}

\subsection{System Architecture and API}

As a distributed similarity search system, Pyramid consists of three kinds of major components, i.e., coordinators, executors and brokers. We plot the architecture of Pyramid in Figure~\ref{fig:sys_arch}. The coordinators receive queries from some upstream applications (e.g, product recommendation and image search), search the meta-HNSW with the queries and send query processing requests to the executors with the relevant sub-HNSWs. The executors conduct search on their own sub-HNSWs with the received query processing requests and return the partial results to the coordinators. The coordinators merge these partial results to obtain the final results. The brokers handle the delivery of the query processing requests from the coordinators to the executors and use Kafka for reliable message passing on unreliable network. A Zookeeper cluster is used to monitor the workers in the system to detect failures. The Kafka brokers also internally rely on Zookeeper.  
 
Pyramid provides a set of high-level API for query processing and index building, which hides the details of distributed processing from users. The \textit{coordinator} and \textit{executor} are two classes central to query processing while the \textit{GraphConstructor} class is used for index building. We introduce the three classes as follows.
      
\vspace{2mm}       
\noindent \textbf{Coordinator.} \ Listing~\ref{lst:coor_api} shows a summary of the coordinator class. To construct a~\texttt{Coordinator} object, a user passes the broker list, the path to the meta-HNSW, the dataset name and the similarity metric definition to the constructor. After that, the~\texttt{execute(query, para)} method could be called to process a query. Alternatively, the \texttt{execute\_async(query, para, callback)} method could be used to asynchronously execute a query. It immediately exits without waiting for the final results and the callback will be invoked when the final results are available.~\texttt{para} provides parameters for query processing, including the branching factor $K$ and the number of required neighbors $k$. Typically, a user writes a program which receives queries from upstream applications using custom protocols (e.g., RESTful API~\cite{fielding2000architectural}), injects the queries to the system through the coordinator API, and returns the results back to the upstream applications.

\begin{lstlisting}[language=C++, caption=API of Coordinator, captionpos=b, label=lst:coor_api]
class Coordinator {
  public:
    Coordinator(brokers, graph_path, name, metric);
    TopK execute(query, para);
    void execute_async(query, para, callback);
};
\end{lstlisting}

\vspace{2mm}       
\noindent \textbf{Executor.} \ Listing~\ref{lst:exec_api} shows a summary of the executor class. To construct an executor, the programmer passes the broker list, the path and ID to a sub-HNSW, the dataset name and the similarity function definition to the constructor. After that, the~\texttt{start(para)} method could be called to start handle the query processing requests, in which~\texttt{para} provides the parameters for query processing (e.g., the search factor $l$ for the bottom layer graph search and the maximum number of similarity function computations for a query). Unlike the coordinators which receive queries from upstream applications through custom protocols, the executors typically do not involve any custom logic. Therefore, a standalone program is provided to directly run an executor without any programming effort.

\begin{lstlisting}[language=C++, caption=API of Executor, captionpos=b, label=lst:exec_api]
class Executor {
  public:
    Executor(brokers, graph_path, name, metric);
    void start(para);
};
\end{lstlisting}

\vspace{2mm}       
\noindent \textbf{GraphConstructor.} \ Listing~\ref{lst:constructor_api} summarizes the GraphConstructor class, which is used to build the index for Pyramid. It takes the dataset path and the similarity metric as input and output the meta-HNSW and sub-HNSWs. The~\texttt{para} provides the parameters for index construction, such as the size of the meta-HNSW $m$ and the number of sub-HNSWs $w$. The~\texttt{refresh()} method reads the dataset again, reconstructs the graphs and notifies the coordinators and executors, which is useful for updating the index when there are changes in the dataset. 

\begin{lstlisting}[language=C++, caption=API of Executor, captionpos=b, label=lst:constructor_api]
class GraphConstructor {
  public:
    GraphConstructor(brokers, dataset_path, 
                     metric, para);
    void refresh();
};
\end{lstlisting}

\subsection{Straggler Mitigation and Fault Tolerance}\label{ssec:impl}

Stragglers and failures are common in large-scale distributed systems. Pyramid relies on replication (which means that the same sub-HNSW is replicated on multiple workers) and Kafka to achieve robustness against straggler and failure. 

\vspace{2mm}       
\noindent \textbf{Straggler Mitigation.} \ The coordinators push query processing requests to the executors through the Kafka brokers and each sub-HNSW forms a Kafka topic. The executors serving the same sub-HNSW form a group and all of them subscribe to the corresponding topic. Stragglers are handled automatically by the message distribution mechanism of Kafka, which periodically re-balances the message queues of the executors. Therefore, the workload of a slow executor is reduced because it receives fewer query processing requests and the requests are offloaded to other executors serving the same sub-HNSW. Beside straggler mitigation, this design also supports elastic scalability, which means that the executors serving the same sub-HNSW can be dynamically added to/removed from the system. In this case, Kafka simply redistributes the message queues among the executors. Currently, Pyramid does not handle straggling coordinators. We assume that the upstream applications use a mechanism (e.g., hashing) to evenly distribute the workload among the coordinators. Moreover, the workload of the coordinators is much lighter comparing with the executors and the influence of straggler may not be significant.          

\vspace{2mm}       
\noindent \textbf{Failure Recovery.} \ There are two kinds of failures in Pyramid, i.e.,~\textit{coordinator failure} and~\textit{executor failure}. Pyramid does not handle the on-going queries when a coordinator fails and relies on the upstream applications to retry another coordinator upon timeout. Note that executors return the partial results to the respective coordinators through bare network connection instead of using Kafka brokers. Therefore, a coordinator serving a retried query only needs to redo all jobs (generating the query processing requests and sending to the executors). If the partial results are sent through Kafka, the new coordinator needs to handle partial states and the retry procedure will be complicated. If there are multiple executors serving the same sub-HNSW and one of them fails, the query processing requests for this sub-HNSW will be handled by the other executors due to the message dispatching mechanism of Kafka. 

Pyramid uses Zookeeper to track the states of the system and a Master monitors the states kept on Zookeeper. On Zookeeper, each running instance (coordinator or executor) should lock a corresponding file. Once the Master finds that a file is unlocked, it restarts the corresponding instance on an available machine. The new instance locks the corresponding file and starts serving. As the failed instance may recover by itself, the new instance exits immediately when it finds that the corresponding file is locked. If the failed instance recovers after the new instance takes up its responsibility (by locking the file), it simply exits. To avoid the single point of failure of the Master, there are several hot backups of Master. A Master serves only if it could lock a file on Zookeeper. The hot backups monitor the file and take up the responsibility of the original Master if the file is unlocked.

\section{Experimental Results}

In this section, we evaluate the performance Pyramid with extensive experiments. First, we explore the influence of the parameters on the performance of Pyramid. Second, we compare Pyramid with two distributed similarity search solutions, HNSW-naive and FLANN~\cite{muja2014flann}. Finally, we evaluate the performance of Pyramid under straggler and failure.
   
\subsection{Performance Metrics and Experimental Setting}

We mainly use three performance metrics in our evaluation, i.e., \textit{precision, throughput and latency}. Their definitions are given as follows.

\vspace{2mm}       
\noindent \textbf{Precision.} \ For top-$k$ similarity search, an algorithm is allowed to return $k$ items. If $k'$ of these $k$ items belong to the ground-truth top-$k$ neighbors of a query, the precision is said to be $k'/k$. Precision measures the quality of similarity search results and higher precision means better quality. Note that precision is a more stringent metric than the \textit{Recall@k} used in~\cite{johnson2017faiss, douze2018link}, which is 1 if the $k$ items returned by an algorithm contains the top-1 neighbor of a query.

\vspace{2mm}       
\noindent \textbf{Throughput and Latency.} \ Query processing throughput is the number of queries that can be processed by a system per second. Query processing latency is the interval between the time a system receives a query and the time the system returns the similarity search results. 

\vspace{2mm} 

Unless otherwise stated, we searched for the top-10 neighbors of a query in the experiments. The reported precision is the average of 10,000 randomly selected queries. For latency, we report the 90th percentile instead of the average, which models the worst-case performance of the system and is more critical to online applications. For the meta-HNSW and the sub-HNSWs in Pyramid, we set their parameters according to the recommendation in the HNSW paper~\cite{malkov2018hnsw} and did not fine-tune them. Specially, the maximum out-degree of an item was set to 32 for the bottom layer graph, while the maximum out-degree was set to 16 for the other layers. The search factor $l$ for graph walk on the bottom layer was set to 100.

We used the three datasets listed in Table~\ref{tab:dataset} for our experiments.  Deep500M was sampled from the Deep1B dataset and contains descriptors of images generated by the last fully connected layer of a deep neural network~\cite{babenko2016efficient}. The SIFT500M dataset was sampled from the SIFT1B dataset and contains 128-dimension SIFT descriptors of images~\cite{jegou2011searching}. The Tiny10M dataset was sampled from the Tiny80M dataset~\footnote{https://groups.csail.mit.edu/vision/TinyImages/} and contains the GIST descriptors of the Tiny images. Both Deep500M and SIFT500M contain items with very similar Euclidean norm, conducting MIPS on them is not interesting as it is equivalent to Euclidean NNS. In contrast, Tiny10M has a wide spread in Euclidean norm distribution and is more suitable for testing the performance of MIPS. Therefore, we conducted MIPS on Tiny10M and Euclidean NNS on Deep500M and SIFT500M.   

All experiments were conducted on a cluster of 10 machines connected with 10 Gbps Ethernet. Each machine is equipped with two 8-core Intel Xeon E5-2620v4 2.1GHz processors and 48GB RAM, running on CentOS 7.2. As the cluster contains 10 machines, we use 10 sub-HNSWs in the experiments.  

\begin{table}[t]
\caption{Datasets}
\begin{center}
\begin{tabular}{|c|c|c|c|}
\hline
\textbf{Name} & \textbf{\# item} & \textbf{\# dimension} & \textbf{size (GB)}\\
\hline
Deep500M		 & 500,000,000     & 96                    &  192\\
\hline
SIFT500M       & 500,000,000     & 128                   &  256\\
\hline
Tiny10M      & 10,000,000        & 384                   &  15.4\\
\hline
\end{tabular}
\label{tab:dataset}
\end{center}
\end{table}

\subsection{Influence of The Parameters} 

\begin{figure}[!t]
	\centering
	\includegraphics[width=0.48\linewidth]{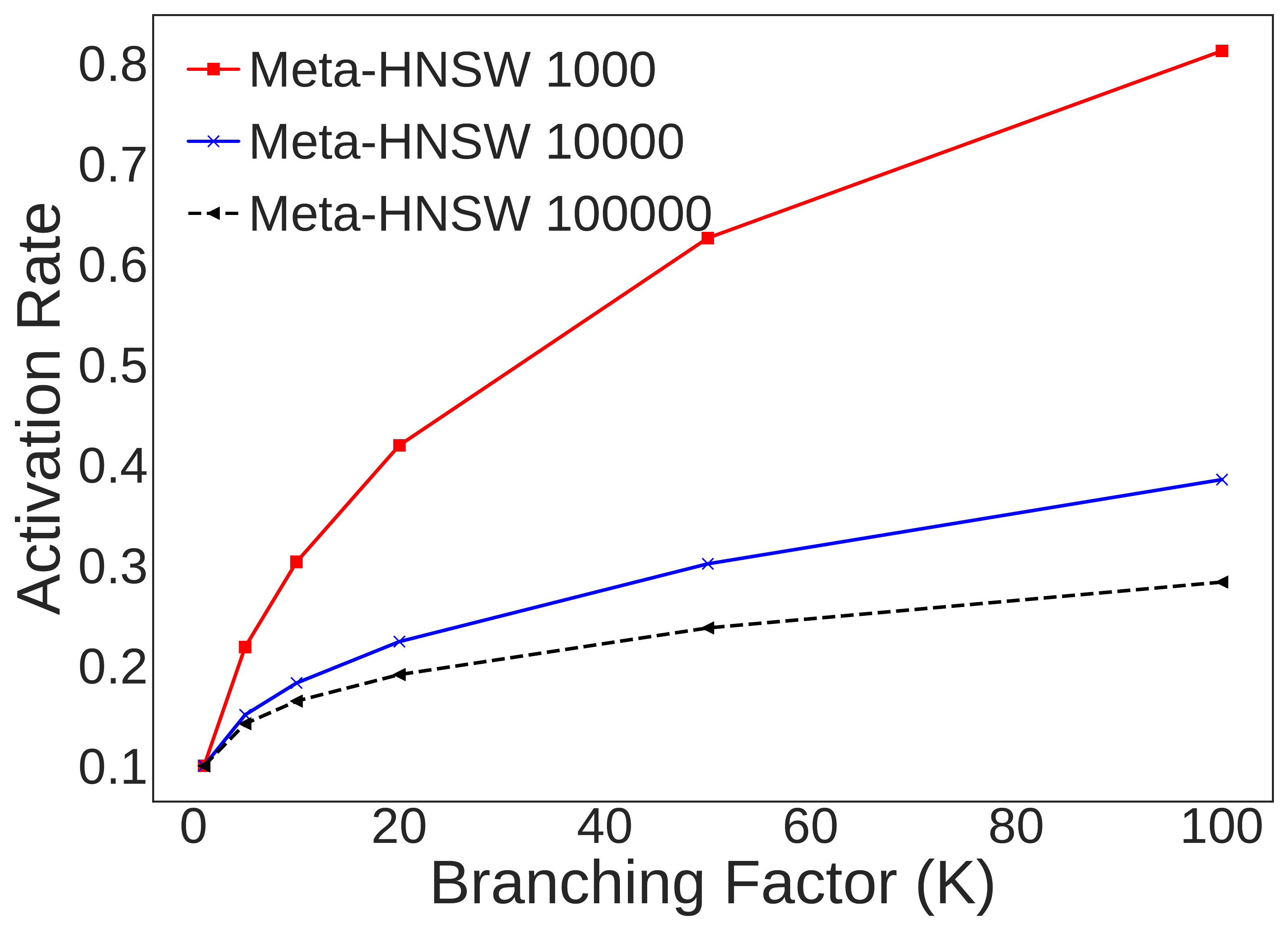}
	\includegraphics[width=0.48\linewidth]{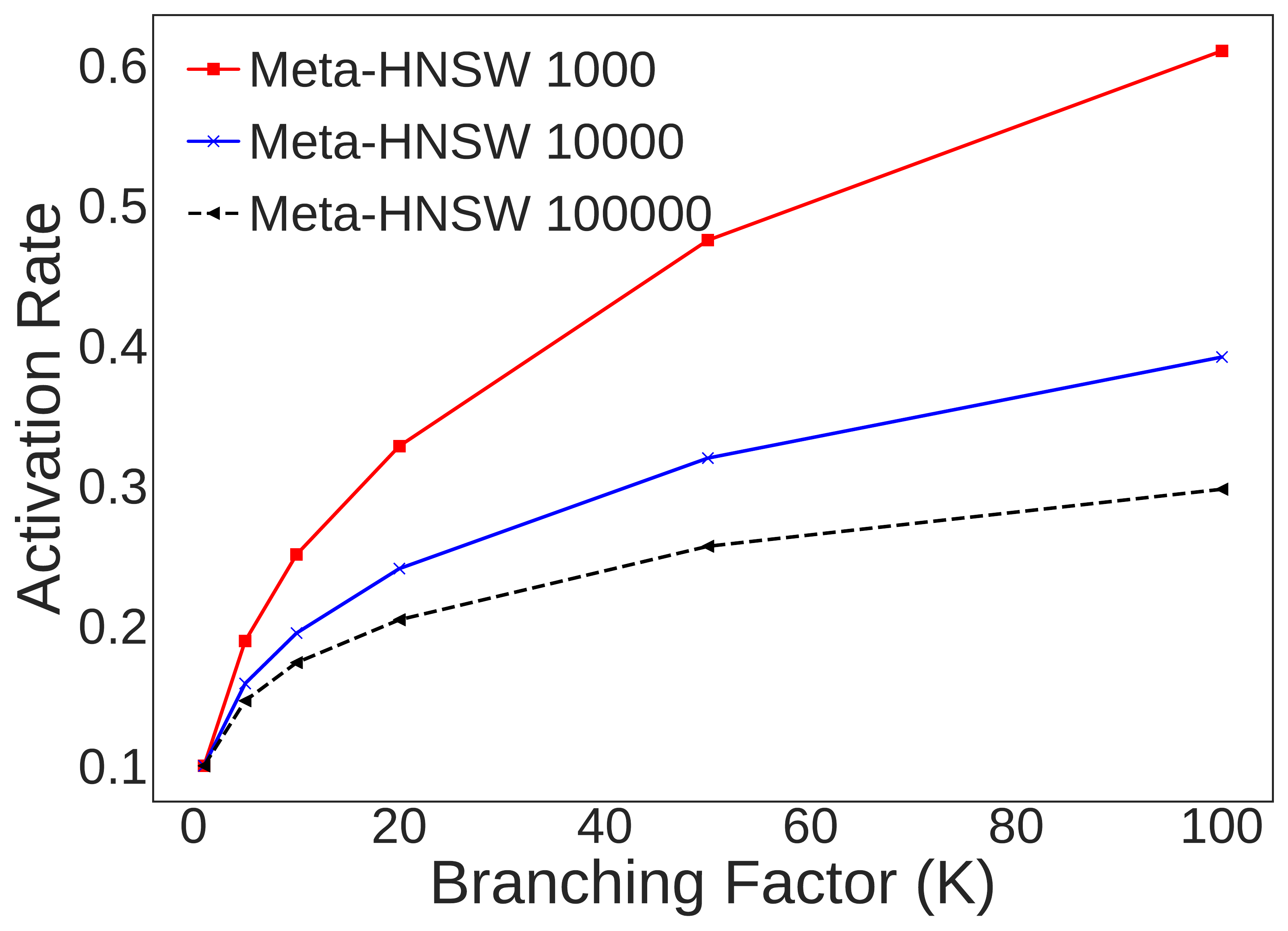}
	\caption{Access rate vs. branching factor for Deep (left) and SIFT (right)}
	\label{fig:access}
\end{figure}

\begin{figure}[!t]
	\centering
	\includegraphics[width=0.48\linewidth]{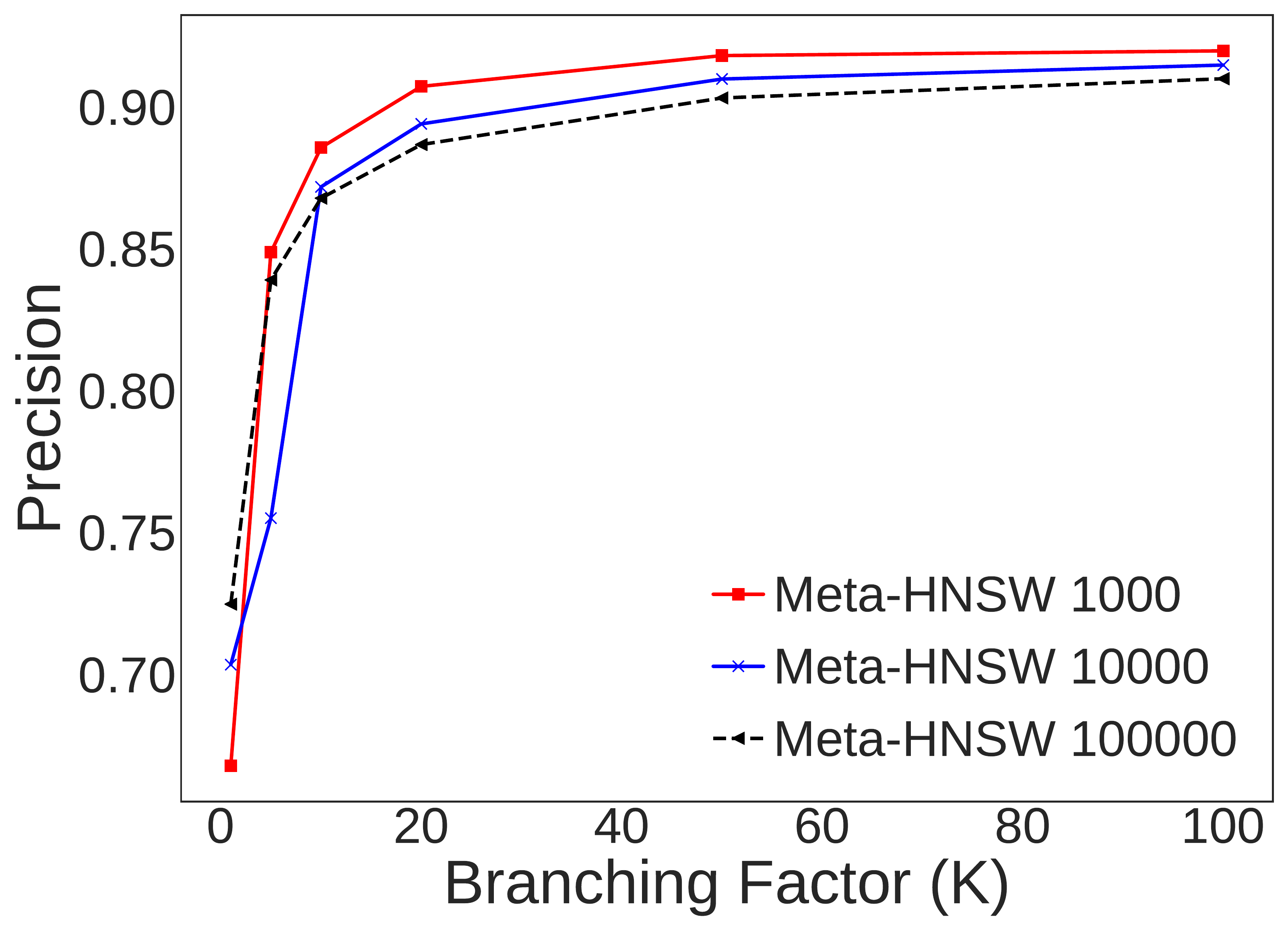}
	\includegraphics[width=0.48\linewidth]{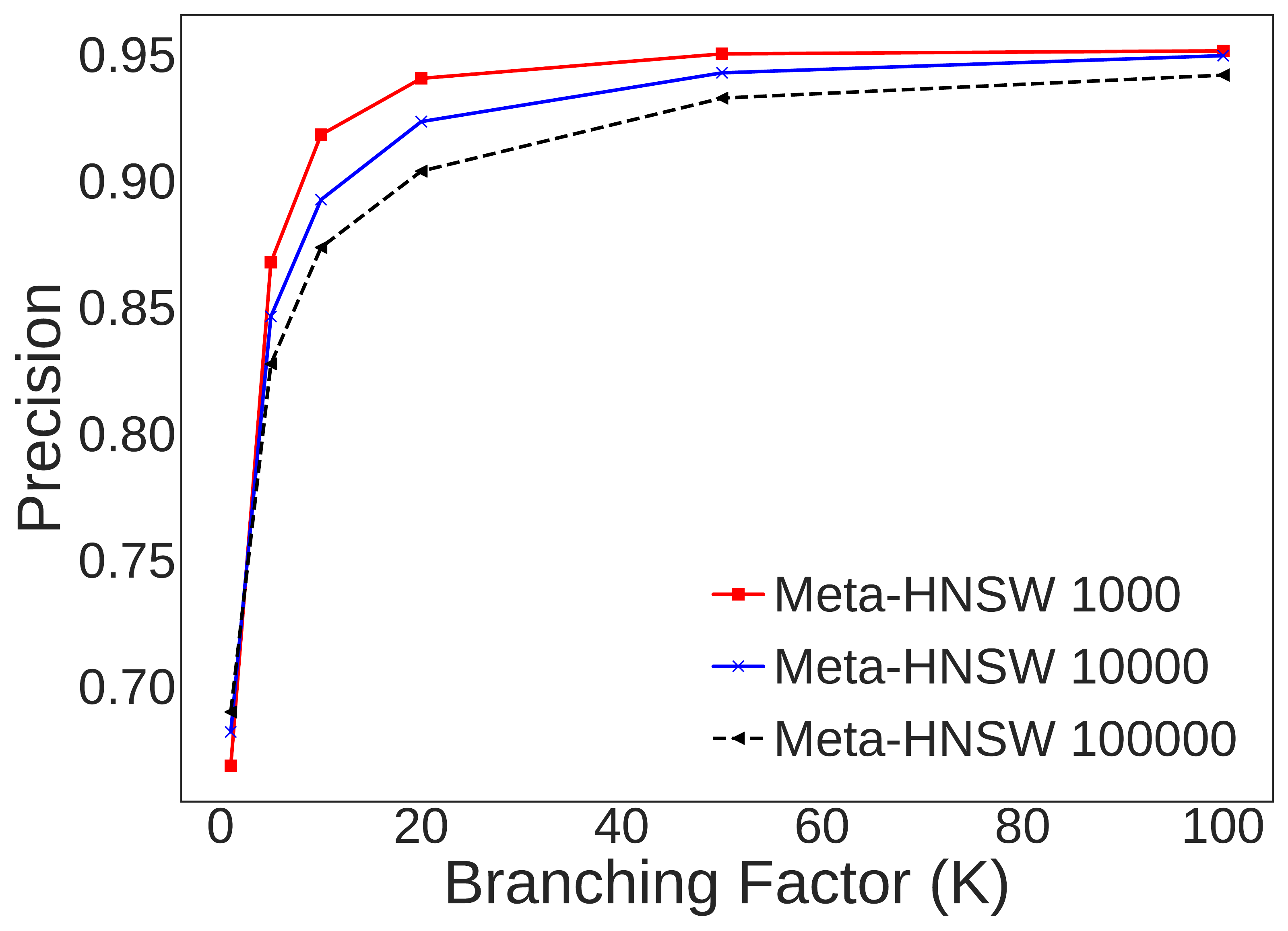}
	\caption{Precision vs. branching factor for Deep (left) and SIFT (right)}
	\label{fig:precision}
\end{figure}

\begin{figure}[!t]
	\centering
	\includegraphics[width=0.48\linewidth]{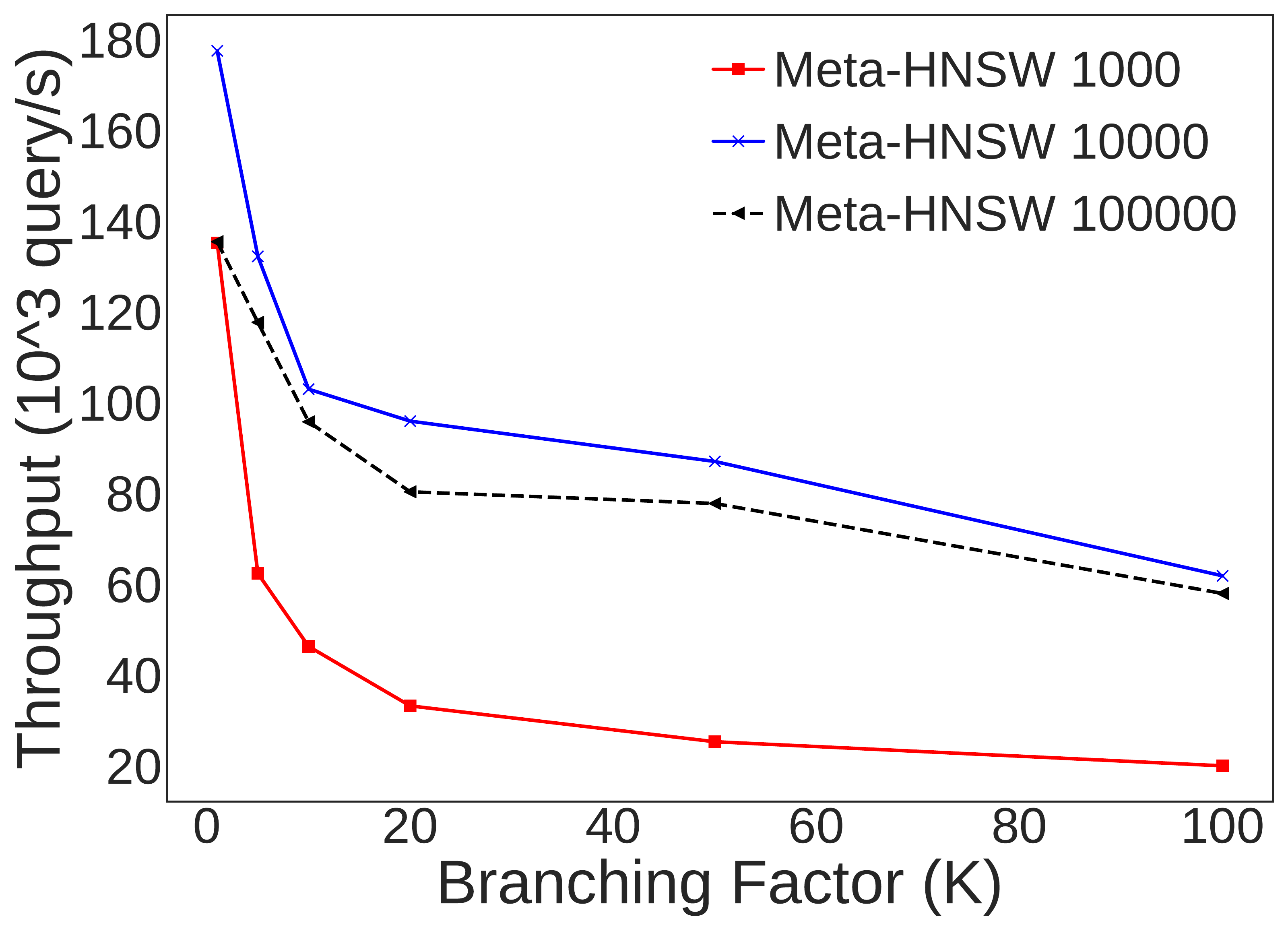}
	\includegraphics[width=0.48\linewidth]{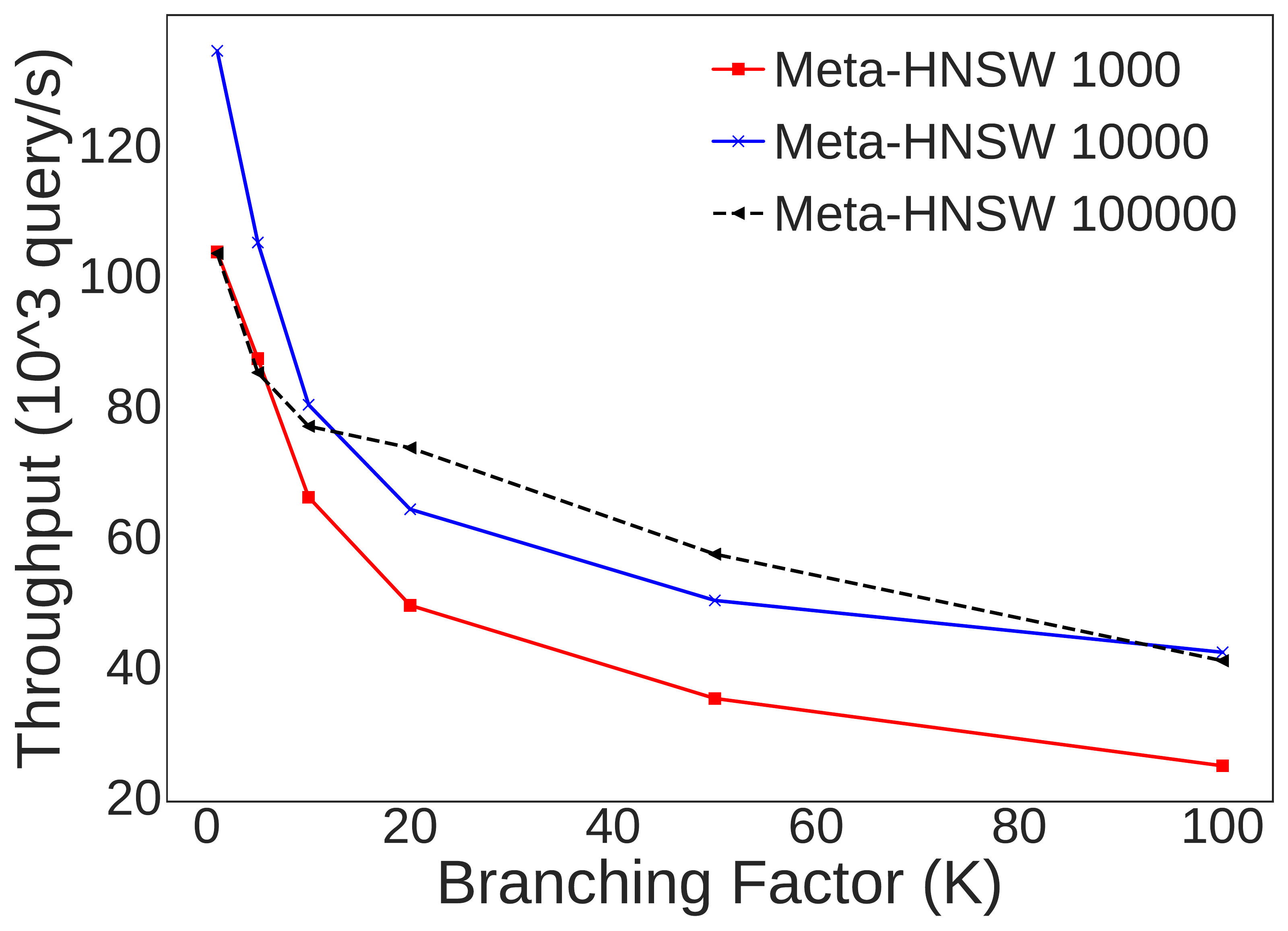}
	\caption{Throughput vs. branching factor for Deep (left) and SIFT (right)}
	\label{fig:throughput}
\end{figure}

\begin{figure}
	\centering
	\includegraphics[width=0.48\linewidth]{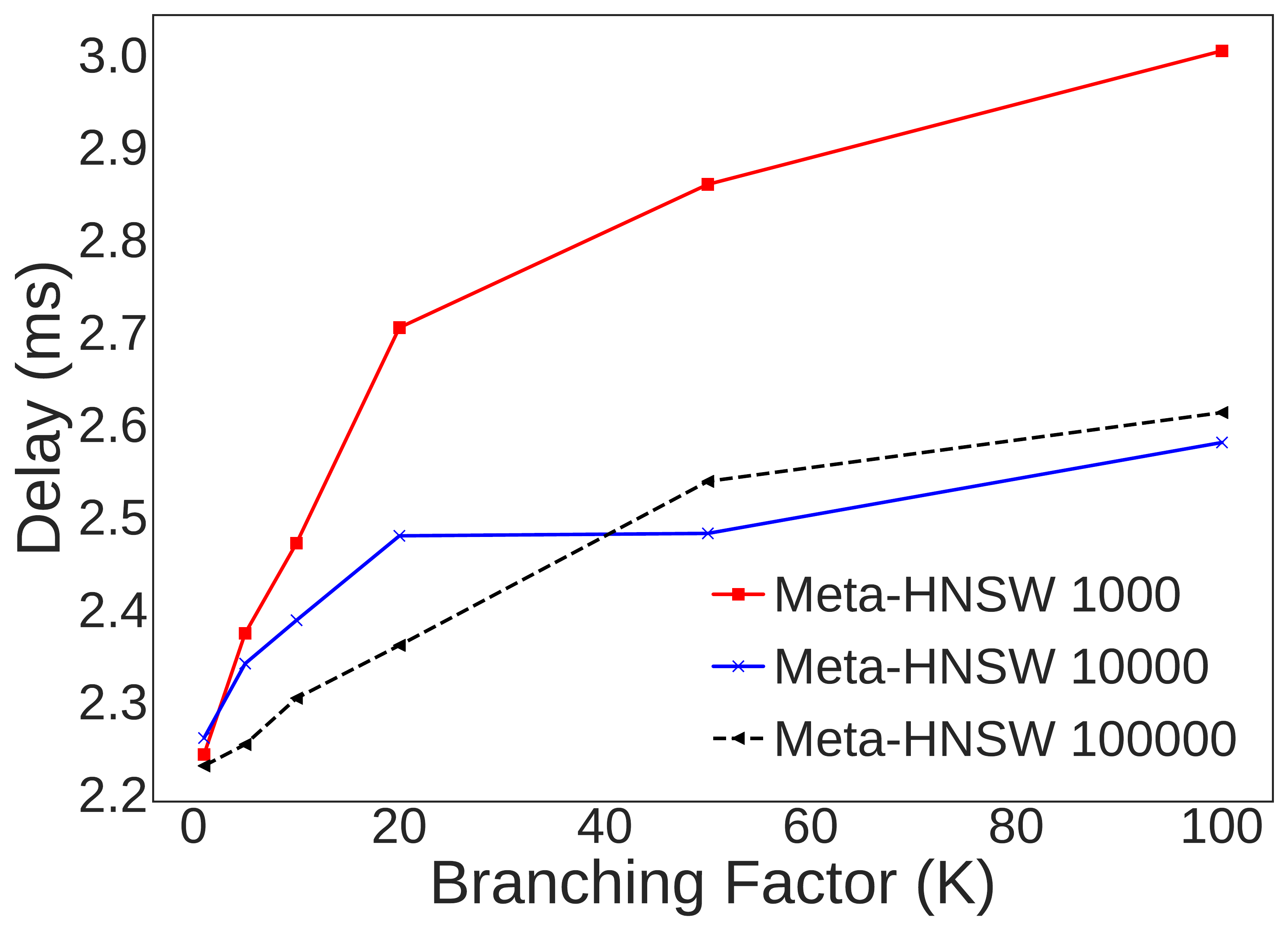}
	\includegraphics[width=0.48\linewidth]{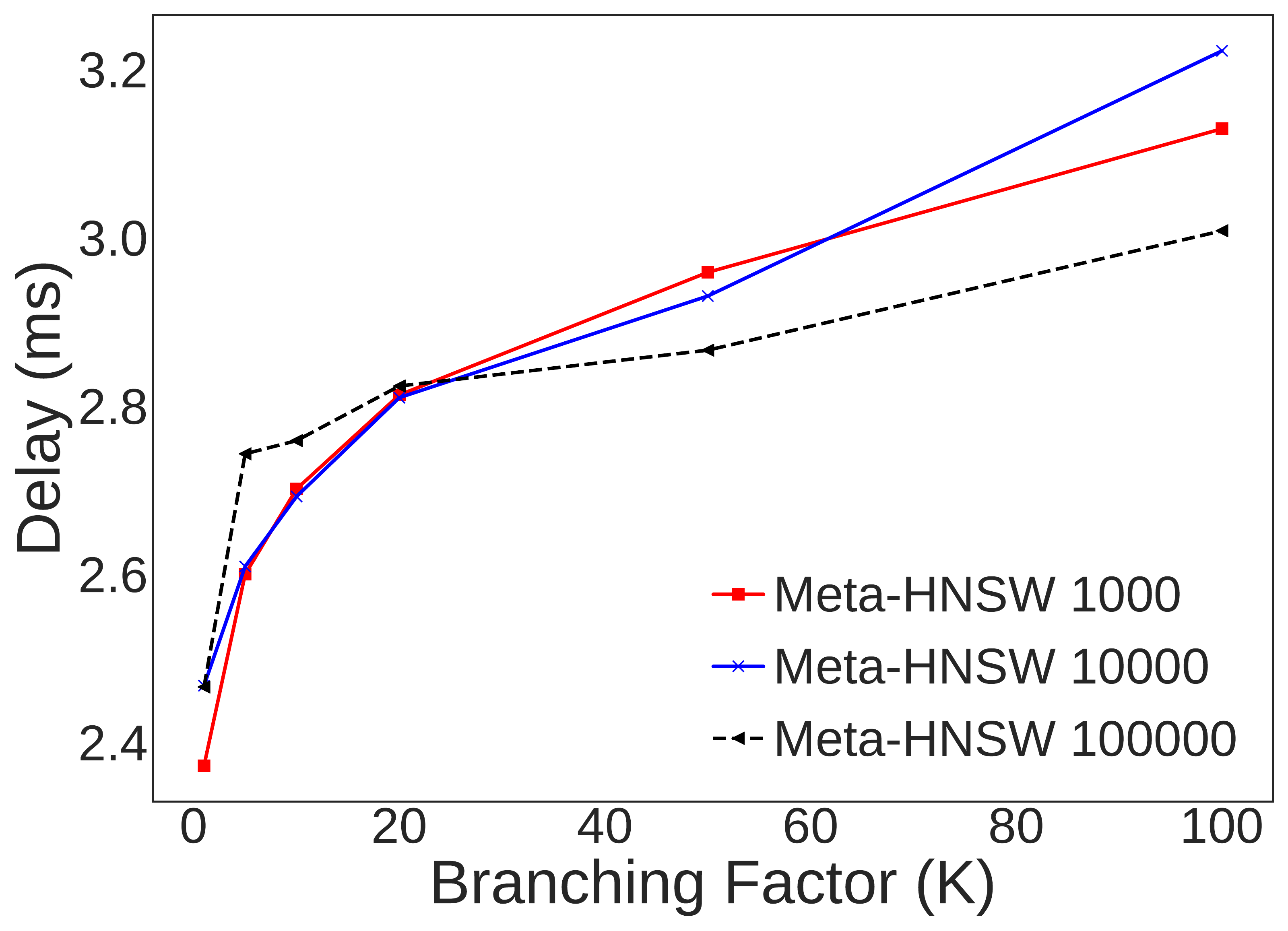}
	\caption{90th PCT latency vs. branching factor for Deep (left) and SIFT (right)}
	\label{fig:latency}
\end{figure}

In this set of experiments, we evaluated the influence of the meta-HNSW related parameters on the performance of Pyramid. Recall that the meta-HNSW has two main parameters, i.e., size $m$ (the number of vectors in the bottom layer graph) and branching factor $K$ (the number of top neighbors in the meta-HNSW that are used to choose the sub-HNSWs for a query). On Deep500M and SIFT500M and for Euclidean NNS, we experimented with a meta-HNSW of  size 1,000, 10,000 and 100,000 and branching factor value of 1, 5, 10, 20, 50 and 100. Different aspects of Pyramid's performance under these configurations are reported in Figure~\ref{fig:access} to Figure~\ref{fig:latency}.   

Figure~\ref{fig:access} reports the average access rate of the queries under different branching factor, where the access rate is the fraction of sub-HNSWs that are accessed for processing a query. Under the same meta-HNSW size, the access rate increases with $K$ as a larger $K$ means that more neighbors in the meta-HNSW are used to choose the sub-HNSWs according to Algorithm~\ref{alg:Pyramid Query}. For the same $K$, a larger meta-HNSW size results in a lower access rate. This is because a larger meta-HNSW offers more fine-grained partitioning of the dataset such that the top $K$ neighbors of a query in the meta-HNSW are contained in a smaller number of sub-datasets.

Figure~\ref{fig:precision} reports the precision of the search results under different configurations. The precision first increases rapidly with the branching factor and then stabilizes. Moreover, the precision is higher for a smaller meta-HNSW size under the same branching factor. This is because more sub-HNSWs are accessed to process a query for a smaller meta-HNSW according to Figure~\ref{fig:access}. Combining Figure~\ref{fig:access} and Figure~\ref{fig:precision}, we can also conclude that Pyramid provides high quality search results with a low sub-HNSW access rate. For example, the precisions on both Deep500M and SIFT500M are above 65\% with $K=1$, under which only one sub-HNSW is accessed. This result verifies the effectiveness of the meta-HNSW based dataset partitioning and query assignment strategies in Pyramid.

Figure~\ref{fig:throughput} reports the query processing throughput under different branching factor $K$. The results show that the throughput consistently drops when $K$ increases. This is because a larger $K$ results in a higher access rate, which means that more sub-HNSWs are searched to answer a query and thus the per-query workload is heavier. Although the access rate is lower for a larger meta-HNSW size under the same $K$ according to Figure~\ref{fig:access}, meta-HNSW 100,000 does not always achieve higher throughput than meta-HNSW 10,000. This is because searching a larger meta-HNSW has higher complexity and we observed that meta-HNSW search takes 0.06ms and 0.18ms per query for meta-HNSW 10,000 and meta-HNSW 100,000, respectively. As the difference between the access rates of meta-HNSW 10,000 and meta-HNSW 100,000 is not large, high meta-HNSW search complexity could outweigh the benefits of a lower access rate. This phenomenon also suggests that further increasing the meta-HNSW size beyond 100,000 may degrade the performance due to even higher meta-HNSW search complexity. Therefore, Pyramid does not require a large meta-HNSW to achieve the optimal performance, which is favorable as the meta-HNSW needs to be replicated on every coordinator and a large meta-HNSW will take up a larger amount of memory.

Figure~\ref{fig:latency} reports the 90th percentile of the query processing latency under different configurations. The results show that the latency increases with the branching factor $K$. This is because a coordinator needs to aggregate the results from more executors due to the large access rate under a large $K$ and the latency depends on the maximum latency in these executors. The influence of the meta-HNSW size on latency is more complicated as a large meta-HNSW reduces the access rate but requires longer time for searching the meta-HNSW.

In conclusion, this set of experiments shows that distributed similarity search with Pyramid can provide high quality results (achieving a precision above 90\%), support high throughput (over 100,000 queries per second) and achieve low latency (2 to 3 milliseconds). As Pyramid performs well with a meta-HNSW size of 10,000 on the two datasets we experimented with, we use a  meta-HNSW size of 10,000 in all the subsequent experiments.

\subsection{Comparison with Other Methods}


In this set of experiments, we compared Pyramid with two other distributed similarity search solutions, i.e., HNSW-naive and FLANN~\cite{muja2014flann}. We are aware that there are other distributed similarity search solutions, such as PLSH~\cite{sundaram2013streaming} and SPTAG~\cite{ChenW18}. We did not compare with them because PLSH is not open-source and SPTAG replicates the entire dataset on every machine and cannot support large datasets. As introduced in Section~\ref{sec:design}, HNSW-naive randomly partitions a dataset among the workers and builds a sub-HNSW on each worker. Therefore, a query needs to be handled by all the workers. FLANN is a widely used library for similarity search and supports distributed similarity search with KD tree. Similar to HNSW-naive, FLANN randomly partitions the data among workers and builds an index on each worker. HNSW-naive and Pyramid used the same maximum out-degree configuration for the sub-HNSWs. To enable a fair throughput comparison between Pyramid and HNSW-naive, we adjusted their query processing parameters (branching factor $K$ and search factor $l$ for Pyramid and search factor $l$ for HNSW-naive) to achieve a precision approximately 90\%. As it is difficult for FLANN to achieve a 90\% precision, we report both its precision and throughput under the setting recommended in~\cite{muja2014flann}.

\begin{figure}[!t]
	\centering
	\includegraphics[width=0.48\linewidth]{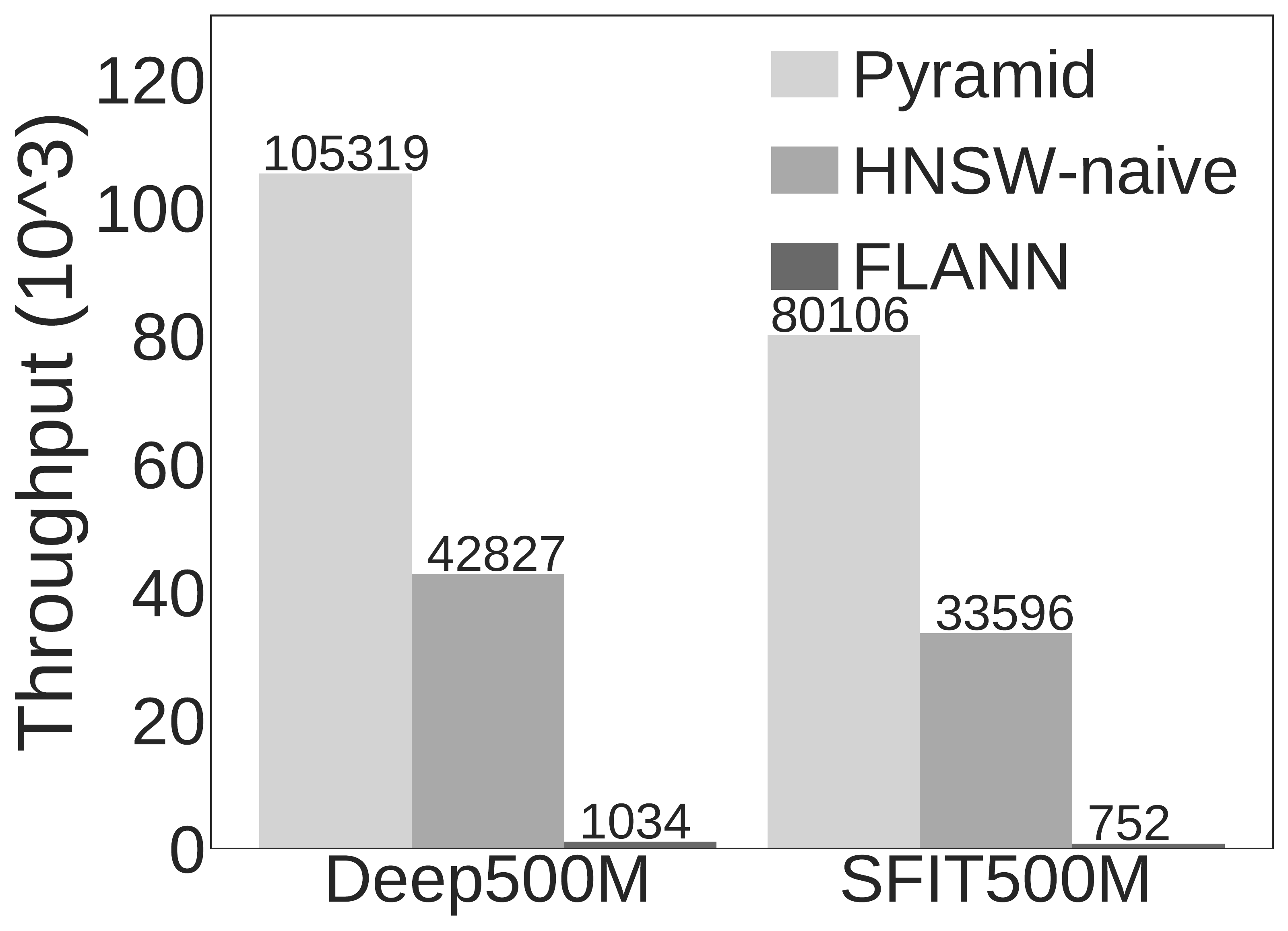}
	\includegraphics[width=0.48\linewidth]{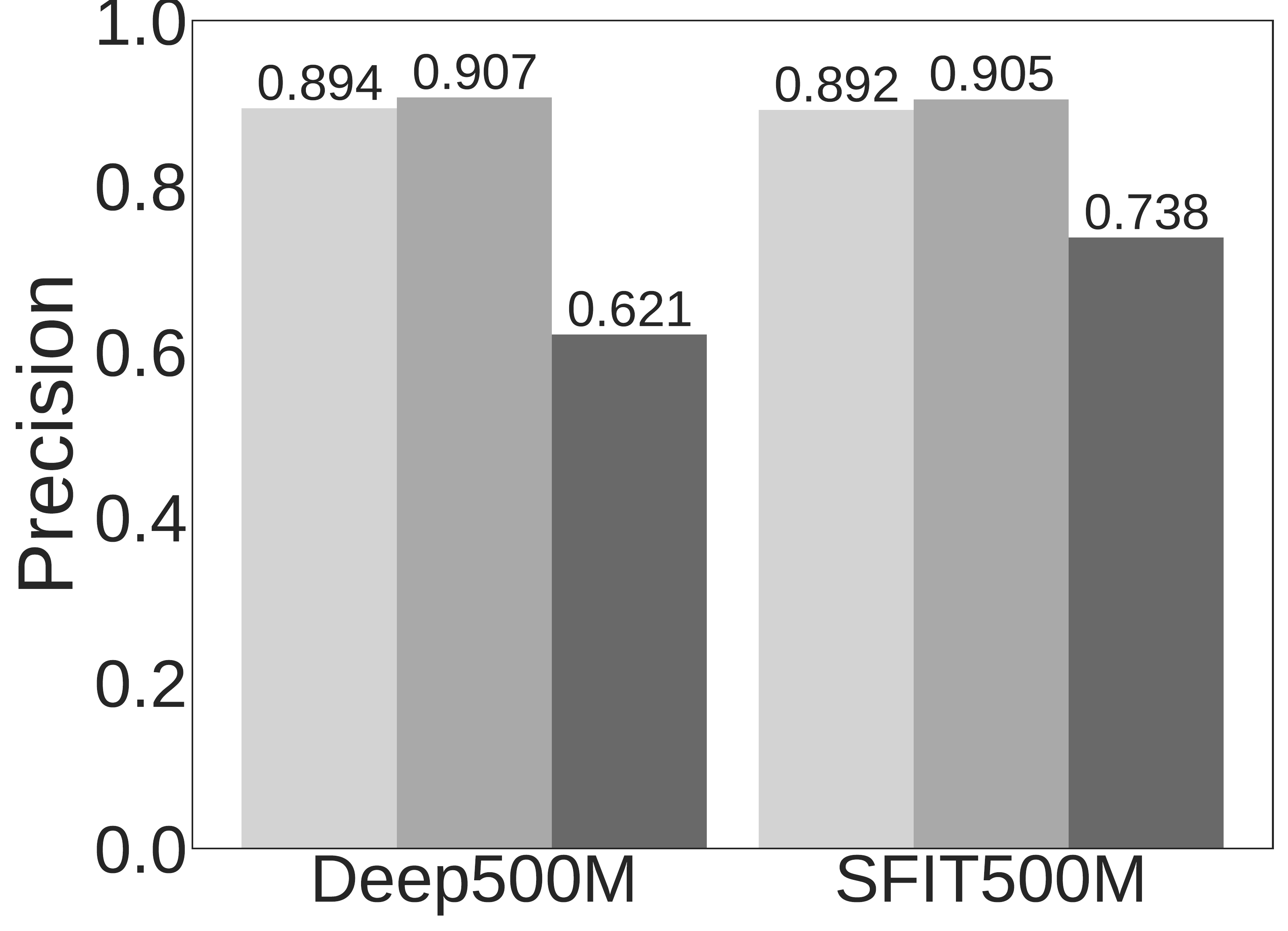}
	\caption{Throughput and precision comparison of the systems}
	\label{fig:compare-deep500M}
\end{figure}

The throughput and precision results of the systems on Deep500M and SIFT500M are reported in Figure~\ref{fig:compare-deep500M}. The results show that Pyramid achieves a throughput that is over 2x of HNSW-naive at a similar precision. This is because Pyramid uses a dataset partitioning and query assignment strategy to avoid searching the  sub-HNSWs in all workers for processing a query. As each query generates less workload in Pyramid, its throughput is much higher than HNSW-naive. Moreover, Pyramid and HNSW-naive outperform FLANN in both throughput and precision due to the algorithmic advantage of HNSW over the tree-based method adopted in FLANN. In particular, the throughput of Pyramid is two orders of magnitude higher than that of FLANN, which shows the benefits of building distributed similarity search solutions based on the state-of-the-art similarity search algorithm.

It is worth noting that the better performance of Pyramid comes at the cost of more expensive index building. For the Deep500M dataset, Pyramid took about 162 minutes for index building using 10 machines, with 31 minutes for meta-HNSW construction, 87 minutes for dataset partitioning and 44 minutes for sub-HNSW construction. As a comparison, HNSW-naive took 53 minutes to build the index, with 14 minutes for dataset partitioning and 39 minutes for sub-HNSW construction. The longer index building time of Pyramid is mainly caused by dataset partitioning using the meta-HNSW, which needs to search the meta-HNSW with every item. For FLANN, index building is extremely fast and took only 38 seconds. As index building is conducted off-line, Pyramid is suitable for applications where dataset updates are not frequent and online search performance is critical.

\begin{figure}[!t]
	\centering
	\includegraphics[width=0.48\linewidth]{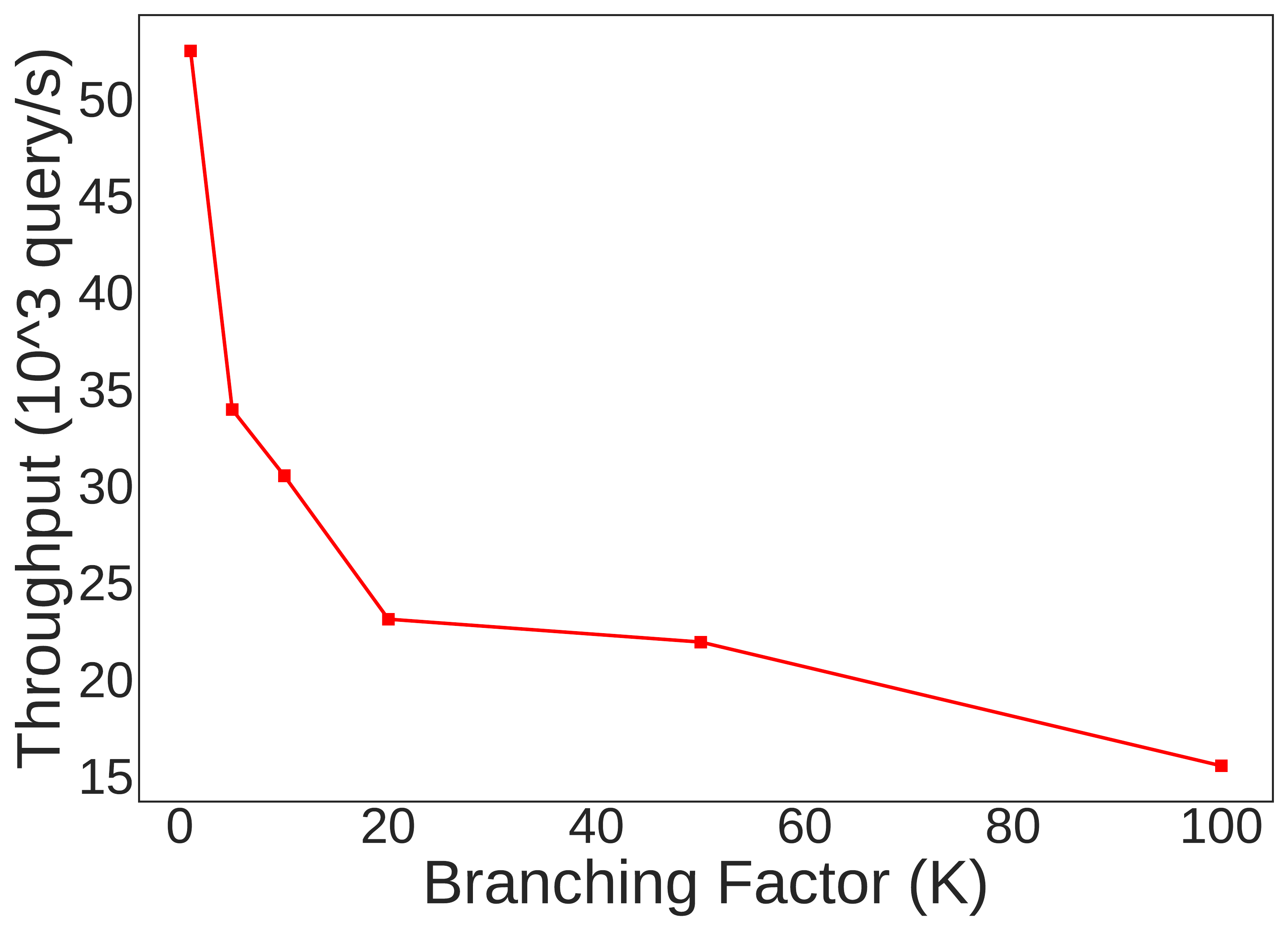}
	\includegraphics[width=0.48\linewidth]{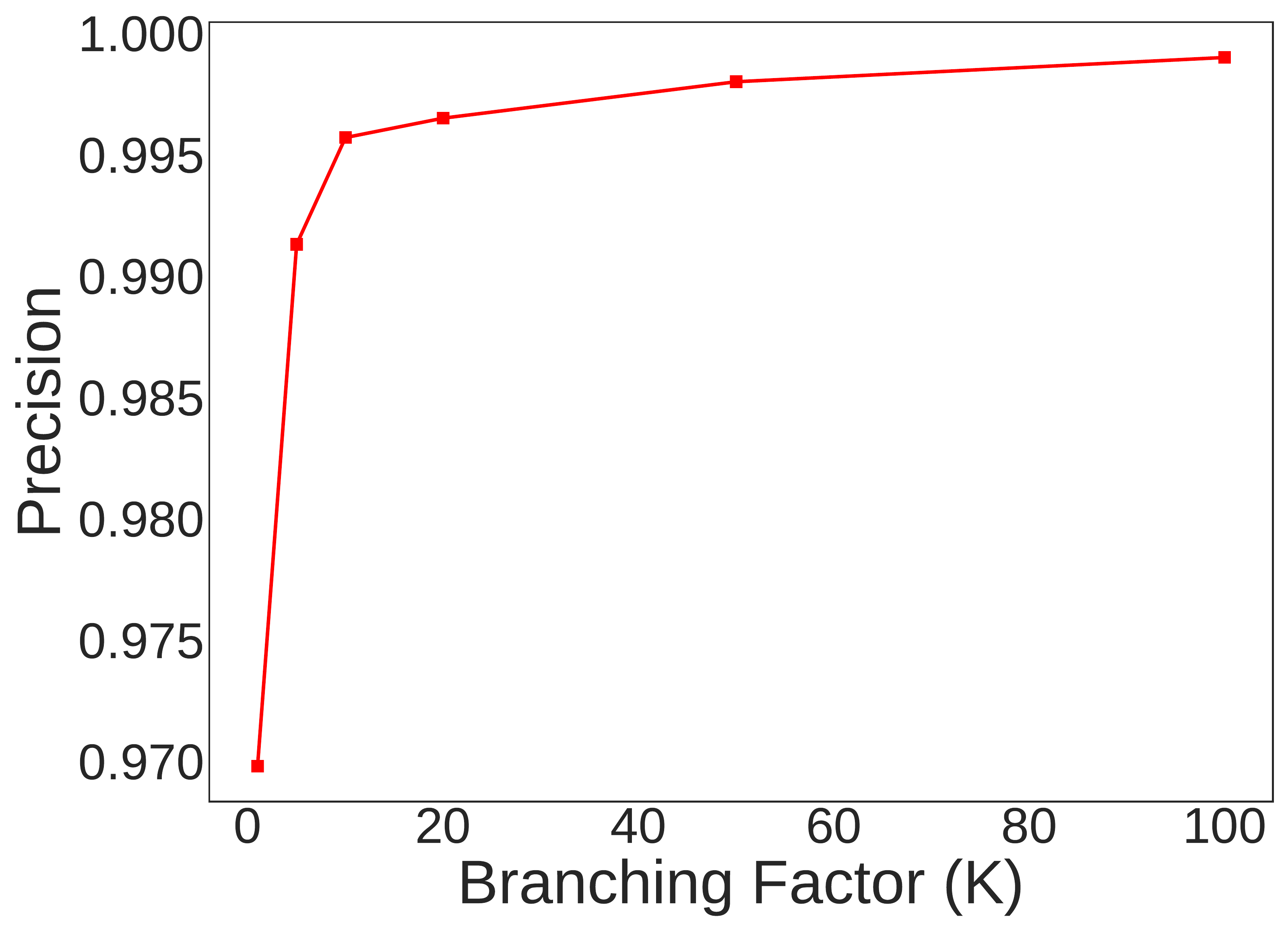}
	\caption{Performance of Pyramid on MIPS}
	\label{fig:compare-MIPS}
\end{figure}

We report the performance of Pyramid on MIPS for the Tiny10M dataset in Figure~\ref{fig:compare-MIPS}. As there are no distributed similarity search solutions that support MIPS, we report the performance of HNSW-naive as the baseline. HNSW-naive achieves a precision of 99.7\% and a throughput of 12,732 queries per second. In comparison, the throughput of Pyramid is much higher under similar precision thanks to its low access rate. Recall that we allow an item to appear in multiple sub-datasets for MIPS in Algorithm~\ref{alg:Pyramid Index MIPS}. Due to this design, Pyramid achieves a precision of 96.98\% under a branching factor of 1, which means that the sub-HNSW in only one of the 10 machines is accessed for processing each query. Note that we did not replicate a lot of items across the machines, which would take up a lot of memory. For this experiment, we set the replication factor $r$ as 300 and all the sub-HNSWs end up storing a total of 10,060,599 items, which is only 0.6\% larger than the original Tiny10M dataset. We believe allowing some items to appear in multiple sub-datasets can be an effective measure to reduce the access rate for distributed similarity search and adopting this idea of Euclidean NNS will be an interesting direction for future study.

\subsection{Scalability, Straggler Mitigation and Fault Tolerance} 

In this set of experiments, we tested the scalability and robustness of Pyramid. The experiments were conducted on the SIFT100M dataset, which was sampled from the SIFT500M dataset.

\begin{figure}[!t]
	\centering
	\includegraphics[width=0.48\linewidth]{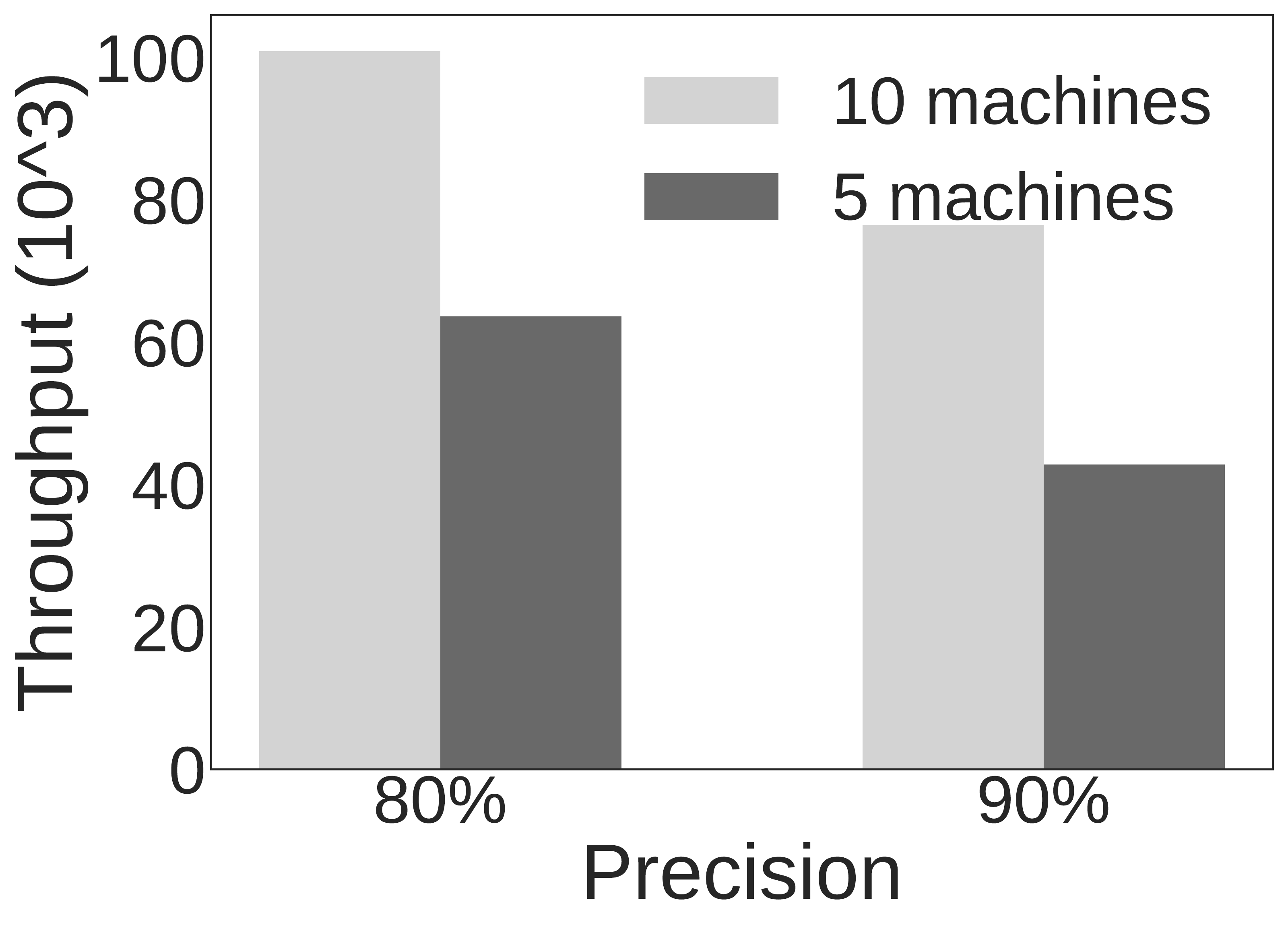}
	\caption{The scalability of Pyramid}
	\label{fig:compare-scaling}
\end{figure}

To check the scalability of Pyramid, we tested its throughput on the SIFT100M dataset with 10 machines and 5 machines. For a fair comparison of throughout, we adjusted the query processing parameters (e.g., branching factor $K$ and search factor $l$) to ensure that both configurations achieved the same precision (80\% and 90\%). We report the results in Figure~\ref{fig:compare-scaling}. The throughput with 10 machines is 1.78 and 1.59 times of the case with 5 machines under a precision of 80\% and 90\%, respectively. The reason that Pyramid does not achieve linear scaling could be due to the characteristic of HNSW. As introduced in Section~\ref{sec:background}, the search complexity of HNSW scales with $O(\log n)$ with $n$ being the cardinality of the dataset. To achieve the same precision, the 5 machine configuration needs to access a smaller number of sub-HNSWs but the number of items in each sub-HNSWs is larger comparing with the 10 machine configuration. As the search complexity of HNSW increases slowly with the dataset cardinality, the 5 machine configuration has a smaller per-query complexity than its 10 machine counterpart because the influence of using a smaller number of sub-HNSWs is dominant.

\begin{figure}[!t]
	\centering
	\includegraphics[width=0.48\linewidth]{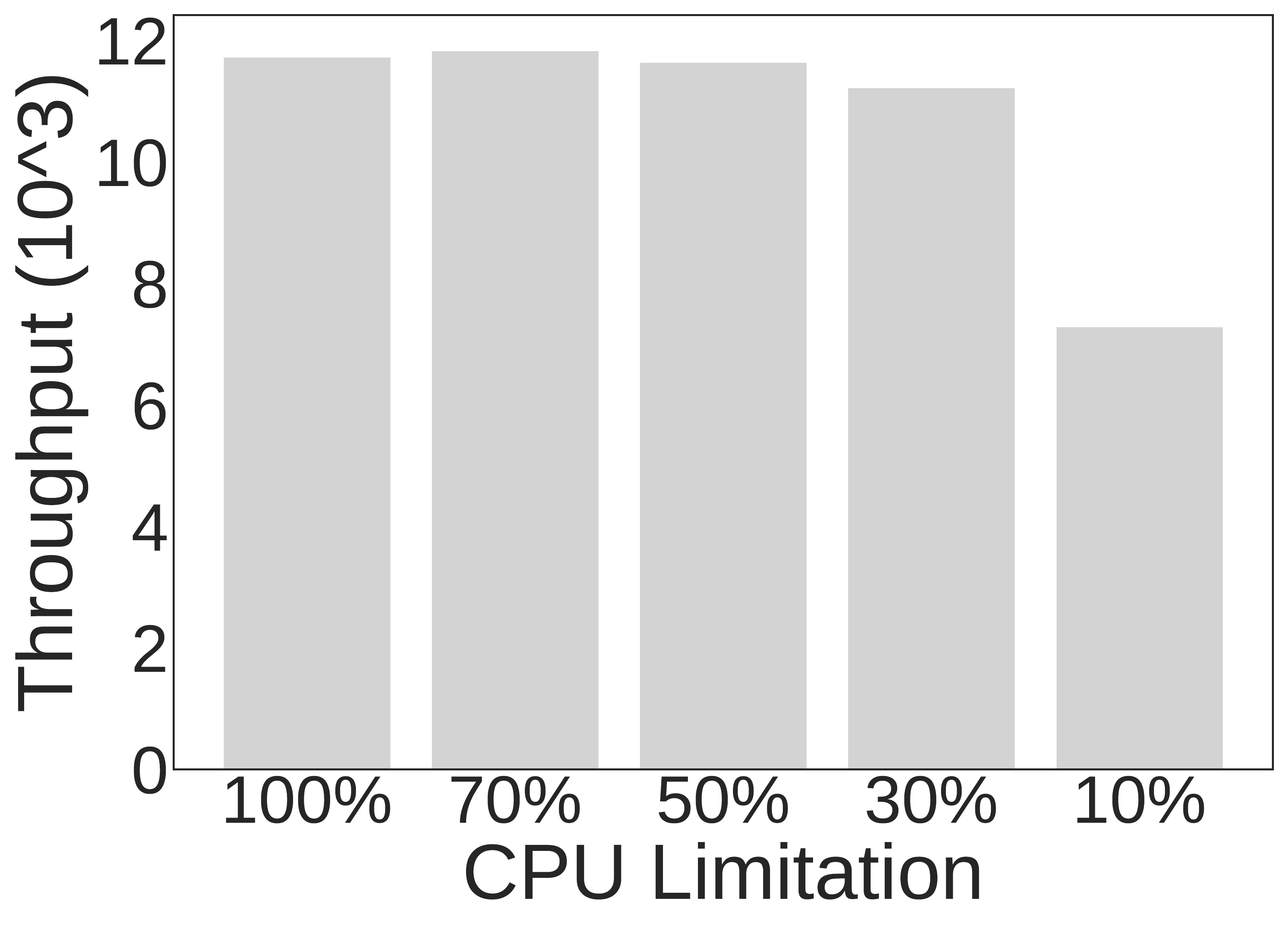}
	\caption{Performance of Pyramid under straggler}
	\label{fig:straggler}
\end{figure}

\begin{figure}[!t]
	\centering
	\includegraphics[width=0.48\linewidth]{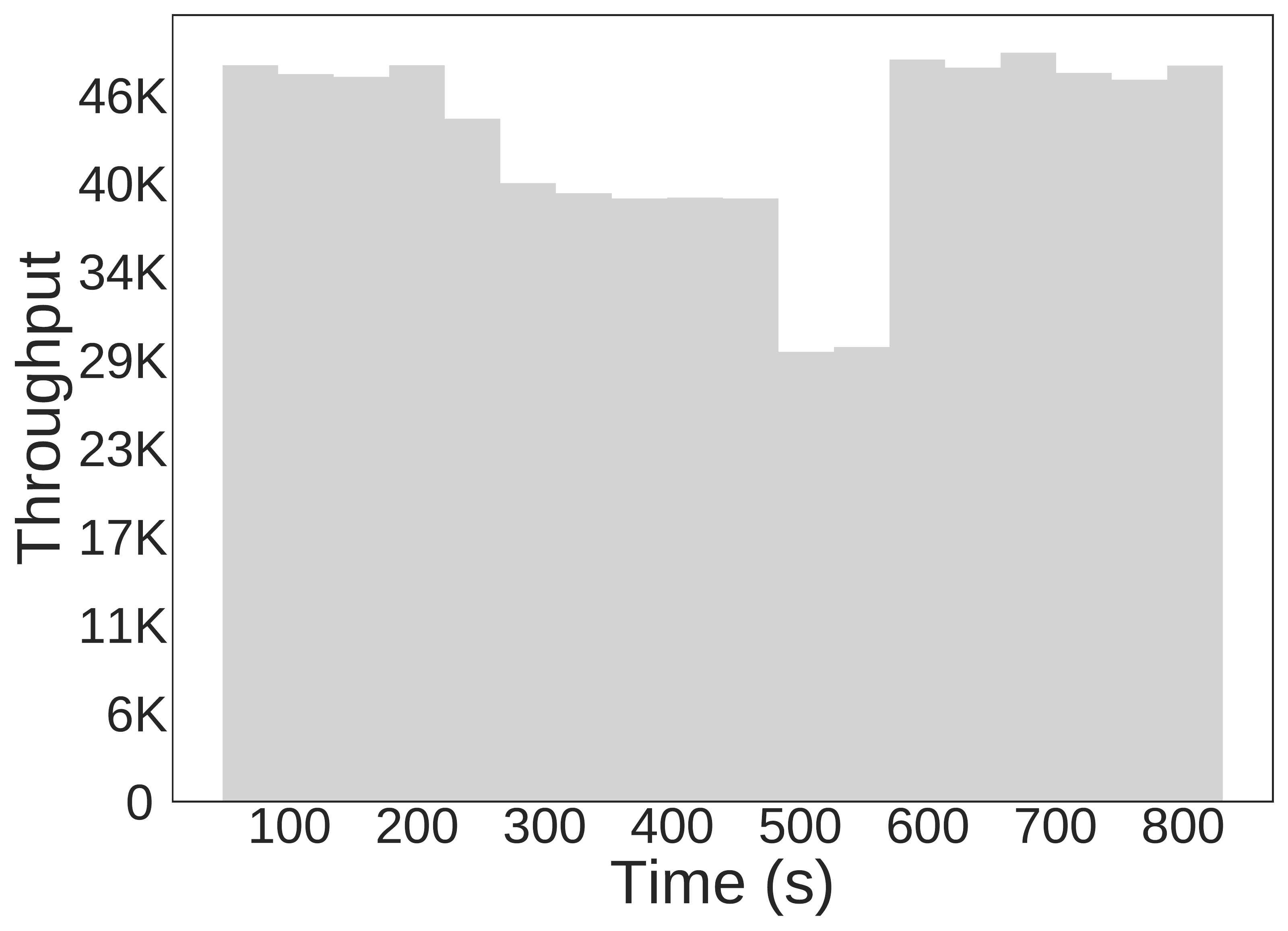}
	\caption{Performance of Pyramid under failure}
	\label{fig:fault}
\end{figure}

To test the performance of Pyramid under straggler, we used the CPU-limit tool to constrain the CPU usage of one of the machines. In this experiment, we created two copies of each sub-HNSW and assigned them to two different machines. Moreover, we also ensured that each machine hosted two different sub-HNSWs. We configured the system to run at 70\% of its peak throughput and measured the throughput of the queries that would access the sub-HNSWs hosted on the CPU limited machine. The results in Figure~\ref{fig:straggler} show that there is no significant change in the throughput when the CPU share is above 30\%. This is because the two sub-HNSWs hosted on the CPU limited machine had replicas on other machines and Kafka would offload the queries to these machines due to its message dispatching mechanism. As the system ran at only 70\% of its peak throughput, the other machines had idle resource to serve the offloaded queries. When the CPU share was too low (e.g., 10\%), the throughput decreases significantly as too many queries were offloaded and the other machines did not have sufficient resource to process them. However, this extreme case of straggler is not common in practice and Pyramid can provide a steady throughput in the case of straggler using replication.   

We report the performance of Pyramid under failure in Figure~\ref{fig:fault}. The results show that the query processing throughput drops at the time when one machine was killed (at around the 300th second). At around the 500th second, when the failed machine rejoined query processing, throughput dropped again because Kafka needed to re-balance the message queue for the machines. When the re-balancing finished at approximately the 600th second, the throughput returned back to the level before failure. The result thus show that Pyramid can effectively handle failure.

\section{Related Work}
\textbf{Algorithms.} Tree-based algorithms, such as KD tree~\cite{fukunage1975branch}, B+ tree~\cite{jagadish2005idistance} and cone tree~\cite{ram2012conetree}, were first proposed for similarity search. However, it was found that these algorithms are prone to curse of dimensionality~\cite{weber1998quantitative}. Locality sensitive hashing (LSH) based methods~\cite{indyk1998approximate, datar2004locality,li2018general} were then proposed to use random hash functions to map the items to buckets and provide theoretical guarantee on the quality of the search results. Recently, the vector quantization (VQ) techniques, such as PQ~\cite{jegou2010pq}, OPQ~\cite{ge2013opq} and CQ~\cite{zhang2014composite}, were proposed, which learn vector codebooks from the data and support both fast similarity function computation and data compression. As the VQ methods are data-dependent, they usually provide better performance than the data-independent LSH-based methods. More recently, the proximity graph based methods~\cite{hajebi2011knngraph, harwood2016fanng, malkov2018hnsw} are shown to significantly outperform other methods. Although similarity search research traditionally focuses on Euclidean NNS, there is an increasing interest in MIPS due to its many applications~\cite{neyshabur2014symmetric, yan2018norm}. It was found that proximity graph also provides the best performance for MIPS~\cite{morozov2018mips}. Pyramid builds its distributed solution based on HNSW, the state of the art proximity graph method, and supports popular similarity functions including Euclidean distance and inner product.

\textbf{Large-scale solutions.} There are a number of works that address similarity search on large-scale datasets for applications such as image search. FAISS uses VQ techniques to compress billion-scale datasets to fit in the memory of GPU and leverages the computation power of GPU to reduce query processing latency~\cite{johnson2017faiss}. Link\&Code uses VQ techniques and a customized interpretation algorithm to compress large datasets to fit in the main memory and builds an HNSW on the compressed vectors. VQ techniques are used for data compression and the IVFADC index structure~\cite{jegou2010pq} is used for candidate generation in~\cite{baranchuk2018revisiting}. These methods are all based on a single machine and the data compression affects the quality of the search results. Moreover, they also have problems in scaling to the large datasets (e.g., with trillions of items) we may encounter in the future.      
\section{Conclusions}

In this paper, we presented Pyramid, a general, efficient and robust framework for distributed similarity search on large datasets. Pyramid was developed based on HNSW, the state-of-the-art algorithm for similarity search. We devised effective data partitioning and query assignment strategies to improve query processing throughput and latency. Pyramid is general and can work with popular similarity functions including Euclidean distance, angular distance and inner product. Pyramid is also robust to straggler and failures due to its system design. Experimental results show that Pyramid provides high quality search results on large datasets and achieves high query processing throughput and low latency.

\bibliography{pyramid}

\begin{thebibliography}{10}

\bibitem{wang2017similaritysurvey}
Jingdong Wang, Ting Zhang, Nicu Sebe, Heng~Tao Shen, et~al.
\newblock A survey on learning to hash.
\newblock {\em IEEE transactions on pattern analysis and machine intelligence},
  40(4):769--790, 2017.

\bibitem{datar2004locality}
Mayur Datar, Nicole Immorlica, Piotr Indyk, and Vahab~S Mirrokni.
\newblock Locality-sensitive hashing scheme based on p-stable distributions.
\newblock In {\em Proceedings of the twentieth annual symposium on
  Computational geometry}, pages 253--262. ACM, 2004.

\bibitem{charikar2002angular}
Moses~S Charikar.
\newblock Similarity estimation techniques from rounding algorithms.
\newblock In {\em Proceedings of the thiry-fourth annual ACM symposium on
  Theory of computing}, pages 380--388. ACM, 2002.

\bibitem{shrivastava2014alsh}
Anshumali Shrivastava and Ping Li.
\newblock Asymmetric lsh (alsh) for sublinear time maximum inner product search
  (mips).
\newblock In {\em Advances in Neural Information Processing Systems}, pages
  2321--2329, 2014.

\bibitem{philbin2007imagesearch}
James Philbin, Ondrej Chum, Michael Isard, Josef Sivic, and Andrew Zisserman.
\newblock Object retrieval with large vocabularies and fast spatial matching.
\newblock In {\em 2007 IEEE Conference on Computer Vision and Pattern
  Recognition}, pages 1--8. IEEE, 2007.

\bibitem{douze2018lowshot}
Matthijs Douze, Arthur Szlam, Bharath Hariharan, and Herv{\'e} J{\'e}gou.
\newblock Low-shot learning with large-scale diffusion.
\newblock In {\em Proceedings of the IEEE Conference on Computer Vision and
  Pattern Recognition}, pages 3349--3358, 2018.

\bibitem{koren2009matrix}
Yehuda Koren, Robert Bell, and Chris Volinsky.
\newblock Matrix factorization techniques for recommender systems.
\newblock {\em Computer}, (8):30--37, 2009.

\bibitem{berlin2015}
Konstantin Berlin, Sergey Koren, Chen-Shan Chin, James~P Drake, Jane~M
  Landolin, and Adam~M Phillippy.
\newblock Assembling large genomes with single-molecule sequencing and
  locality-sensitive hashing.
\newblock {\em Nature biotechnology}, 33(6):623, 2015.

\bibitem{hoffart2012kore}
Johannes Hoffart, Stephan Seufert, Dat~Ba Nguyen, Martin Theobald, and Gerhard
  Weikum.
\newblock Kore: keyphrase overlap relatedness for entity disambiguation.
\newblock In {\em Proceedings of the 21st ACM international conference on
  Information and knowledge management}, pages 545--554. ACM, 2012.

\bibitem{chandar2016memory}
Sarath Chandar, Sungjin Ahn, Hugo Larochelle, Pascal Vincent, Gerald Tesauro,
  and Yoshua Bengio.
\newblock Hierarchical memory networks.
\newblock {\em arXiv preprint arXiv:1605.07427}, 2016.

\bibitem{jun2017reinforcement}
Kwang-Sung Jun, Aniruddha Bhargava, Robert Nowak, and Rebecca Willett.
\newblock Scalable generalized linear bandits: Online computation and hashing.
\newblock In {\em Advances in Neural Information Processing Systems}, pages
  99--109, 2017.

\bibitem{fukunage1975branch}
K~Fukunage and Patrenahalli~M. Narendra.
\newblock A branch and bound algorithm for computing k-nearest neighbors.
\newblock {\em IEEE transactions on computers}, (7):750--753, 1975.

\bibitem{jagadish2005idistance}
Hosagrahar~V Jagadish, Beng~Chin Ooi, Kian-Lee Tan, Cui Yu, and Rui Zhang.
\newblock idistance: An adaptive b+-tree based indexing method for nearest
  neighbor search.
\newblock {\em ACM Transactions on Database Systems (TODS)}, 30(2):364--397,
  2005.

\bibitem{ram2012conetree}
Parikshit Ram and Alexander~G Gray.
\newblock Maximum inner-product search using cone trees.
\newblock In {\em Proceedings of the 18th ACM SIGKDD international conference
  on Knowledge discovery and data mining}, pages 931--939. ACM, 2012.

\bibitem{indyk1998approximate}
Piotr Indyk and Rajeev Motwani.
\newblock Approximate nearest neighbors: towards removing the curse of
  dimensionality.
\newblock In {\em Proceedings of the thirtieth annual ACM symposium on Theory
  of computing}, pages 604--613. ACM, 1998.

\bibitem{neyshabur2014symmetric}
Behnam Neyshabur and Nathan Srebro.
\newblock On symmetric and asymmetric lshs for inner product search.
\newblock {\em arXiv preprint arXiv:1410.5518}, 2014.

\bibitem{jegou2010pq}
Herve Jegou, Matthijs Douze, and Cordelia Schmid.
\newblock Product quantization for nearest neighbor search.
\newblock {\em IEEE transactions on pattern analysis and machine intelligence},
  33(1):117--128, 2010.

\bibitem{ge2013opq}
Tiezheng Ge, Kaiming He, Qifa Ke, and Jian Sun.
\newblock Optimized product quantization for approximate nearest neighbor
  search.
\newblock In {\em Proceedings of the IEEE Conference on Computer Vision and
  Pattern Recognition}, pages 2946--2953, 2013.

\bibitem{babenko2014imi}
Artem Babenko and Victor Lempitsky.
\newblock The inverted multi-index.
\newblock {\em IEEE Transactions on Pattern Analysis and Machine Intelligence},
  37(6):1247--1260, 2014.

\bibitem{hajebi2011knngraph}
Kiana Hajebi, Yasin Abbasi-Yadkori, Hossein Shahbazi, and Hong Zhang.
\newblock Fast approximate nearest-neighbor search with k-nearest neighbor
  graph.
\newblock In {\em Twenty-Second International Joint Conference on Artificial
  Intelligence}, 2011.

\bibitem{harwood2016fanng}
Ben Harwood and Tom Drummond.
\newblock Fanng: Fast approximate nearest neighbour graphs.
\newblock In {\em Proceedings of the IEEE Conference on Computer Vision and
  Pattern Recognition}, pages 5713--5722, 2016.

\bibitem{malkov2018hnsw}
Yury~A Malkov and Dmitry~A Yashunin.
\newblock Efficient and robust approximate nearest neighbor search using
  hierarchical navigable small world graphs.
\newblock {\em IEEE transactions on pattern analysis and machine intelligence},
  2018.

\bibitem{wang2013graphimi}
Jing Wang, Jingdong Wang, Gang Zeng, Rui Gan, Shipeng Li, and Baining Guo.
\newblock Fast neighborhood graph search using cartesian concatenation.
\newblock In {\em Proceedings of the IEEE International Conference on Computer
  Vision}, pages 2128--2135, 2013.

\bibitem{fu2019nsg}
Cong Fu, Chao Xiang, Changxu Wang, and Deng Cai.
\newblock Fast approximate nearest neighbor search with the navigating
  spreading-out graph.
\newblock {\em Proceedings of the VLDB Endowment}, 12(5):461--474, 2019.

\bibitem{muja2014flann}
Marius Muja and David~G Lowe.
\newblock Scalable nearest neighbor algorithms for high dimensional data.
\newblock {\em IEEE transactions on pattern analysis and machine intelligence},
  36(11):2227--2240, 2014.

\bibitem{sundaram2013streaming}
Narayanan Sundaram, Aizana Turmukhametova, Nadathur Satish, Todd Mostak, Piotr
  Indyk, Samuel Madden, and Pradeep Dubey.
\newblock Streaming similarity search over one billion tweets using parallel
  locality-sensitive hashing.
\newblock {\em Proceedings of the VLDB Endowment}, 6(14):1930--1941, 2013.

\bibitem{zhang2016shuffle}
Wanxin Zhang, Dongsheng Li, Ying Xu, and Yiming Zhang.
\newblock Shuffle-efficient distributed locality sensitive hashing on spark.
\newblock In {\em 2016 IEEE Conference on Computer Communications Workshops
  (INFOCOM WKSHPS)}, pages 766--767. IEEE, 2016.

\bibitem{jegou2011sift}
Herv{\'e} J{\'e}gou, Romain Tavenard, Matthijs Douze, and Laurent Amsaleg.
\newblock Searching in one billion vectors: re-rank with source coding.
\newblock In {\em 2011 IEEE International Conference on Acoustics, Speech and
  Signal Processing (ICASSP)}, pages 861--864. IEEE, 2011.

\bibitem{johnson2017faiss}
Jeff Johnson, Matthijs Douze, and Herv{\'e} J{\'e}gou.
\newblock Billion-scale similarity search with gpus.
\newblock {\em arXiv preprint arXiv:1702.08734}, 2017.

\bibitem{douze2018link}
Matthijs Douze, Alexandre Sablayrolles, and Herv{\'e} J{\'e}gou.
\newblock Link and code: Fast indexing with graphs and compact regression
  codes.
\newblock In {\em Proceedings of the IEEE Conference on Computer Vision and
  Pattern Recognition}, pages 3646--3654, 2018.

\bibitem{abuzaid2019maximus}
Firas Abuzaid, Geet Sethi, Peter Bailis, and Matei Zaharia.
\newblock To index or not to index: Optimizing exact maximum inner product
  search.

\bibitem{baranchuk2018revisiting}
Dmitry Baranchuk, Artem Babenko, and Yury Malkov.
\newblock Revisiting the inverted indices for billion-scale approximate nearest
  neighbors.
\newblock In {\em Proceedings of the European Conference on Computer Vision
  (ECCV)}, pages 202--216, 2018.

\bibitem{morozov2018mips}
Stanislav Morozov and Artem Babenko.
\newblock Non-metric similarity graphs for maximum inner product search.
\newblock In {\em Advances in Neural Information Processing Systems}, pages
  4721--4730, 2018.

\bibitem{sanders2012think}
Peter Sanders and Christian Schulz.
\newblock Think locally, act globally: Perfectly balanced graph partitioning.
\newblock {\em arXiv preprint arXiv:1210.0477}, 2012.

\bibitem{auvolat2015clustering}
Alex Auvolat, Sarath Chandar, Pascal Vincent, Hugo Larochelle, and Yoshua
  Bengio.
\newblock Clustering is efficient for approximate maximum inner product search.
\newblock {\em arXiv preprint arXiv:1507.05910}, 2015.

\bibitem{fielding2000architectural}
Roy~T Fielding and Richard~N Taylor.
\newblock {\em Architectural styles and the design of network-based software
  architectures}, volume~7.
\newblock University of California, Irvine Doctoral dissertation, 2000.

\bibitem{babenko2016efficient}
Artem Babenko and Victor Lempitsky.
\newblock Efficient indexing of billion-scale datasets of deep descriptors.
\newblock In {\em Proceedings of the IEEE Conference on Computer Vision and
  Pattern Recognition}, pages 2055--2063, 2016.

\bibitem{jegou2011searching}
Herv{\'e} J{\'e}gou, Romain Tavenard, Matthijs Douze, and Laurent Amsaleg.
\newblock Searching in one billion vectors: re-rank with source coding.
\newblock In {\em 2011 IEEE International Conference on Acoustics, Speech and
  Signal Processing (ICASSP)}, pages 861--864. IEEE, 2011.

\bibitem{ChenW18}
Qi~Chen, Haidong Wang, Mingqin Li, Gang Ren, Scarlett Li, Jeffery Zhu, Jason
  Li, Chuanjie Liu, Lintao Zhang, and Jingdong Wang.
\newblock {\em SPTAG: A library for fast approximate nearest neighbor search},
  2018.

\bibitem{weber1998quantitative}
Roger Weber, Hans-J{\"o}rg Schek, and Stephen Blott.
\newblock A quantitative analysis and performance study for similarity-search
  methods in high-dimensional spaces.
\newblock In {\em VLDB}, volume~98, pages 194--205, 1998.

\bibitem{li2018general}
Jinfeng Li, Xiao Yan, Jian Zhang, An~Xu, James Cheng, Jie Liu, Kelvin~KW Ng,
  and Ti-chung Cheng.
\newblock A general and efficient querying method for learning to hash.
\newblock In {\em Proceedings of the 2018 International Conference on
  Management of Data}, pages 1333--1347. ACM, 2018.

\bibitem{zhang2014composite}
Ting Zhang, Chao Du, and Jingdong Wang.
\newblock Composite quantization for approximate nearest neighbor search.
\newblock In {\em ICML}, volume~2, page~3, 2014.

\bibitem{yan2018norm}
Xiao Yan, Jinfeng Li, Xinyan Dai, Hongzhi Chen, and James Cheng.
\newblock Norm-ranging lsh for maximum inner product search.
\newblock In {\em Advances in Neural Information Processing Systems}, pages
  2952--2961, 2018.

\end{thebibliography}
\bibliographystyle{unsrt}

\end{document}